\documentclass{ar-1col-S2O}

\usepackage[comma]{natbib}
\usepackage{url}
\usepackage{amssymb}

\definecolor{teal}{rgb}{0.0, 0.5, 0.5}
\definecolor{tealgreen}{rgb}{0.0, 0.51, 0.5}
\definecolor{darkgreen}{rgb}{0.0, 0.2, 0.13}

\jname{Xxxx. Xxx. Xxx. Xxx.}
\jvol{AA}
\jyear{YYYY}
\doi{10.1146/((please add article doi))}

\begin{document}

\markboth{Skowron \& Soszyński}{Long Period Variables}

\title{The Role of Long Period Variable Stars in Observational Astrophysics}

\author{Dorota M. Skowron$^1$ and Igor Soszyński$^1$
\affil{$^1$Astronomical Observatory, University of Warsaw,\\ Aleje Ujazdowskie 4, 00-478, Warsaw, Poland;\\email: dszczyg@astrouw.edu.pl}
}

\begin{abstract}
Long-period variables (LPVs) are evolved red giant and supergiant stars whose pulsations provide unique insights into late stages of stellar evolution and serve as essential tools in modern astrophysics. Their period--luminosity and period--age relations make them valuable distance and age indicators, while their light curve morphology, amplitudes, and multiperiodicity reveal the underlying physics of stellar interiors and mass-loss. In this review, we provide an overview of the current status of LPV studies, focusing on their observational properties and applications, including:

\hangindent=.3cm$\bullet$ Modern classification of LPVs into Miras, semiregular variables (SRVs), and OGLE small-amplitude red giants (OSARGs), which occupy multiple period--luminosity sequences associated with different pulsation modes, chemical compositions, and evolutionary stages

\hangindent=.3cm$\bullet$ Mira variables as reliable distance indicators across diverse stellar environments and their increasing role as standard candles

\hangindent=.3cm$\bullet$ The increasing role of SRVs and OSARGs

\hangindent=.3cm$\bullet$ Long secondary period (LSP) variables as potential tracers of exoplanets

Together with advances in theoretical modeling, these developments establish LPVs as valuable tracers of Galactic structure, stellar populations, and the extragalactic distance scale.

\end{abstract}

\maketitle

\tableofcontents

\section{INTRODUCTION}

Long-period variables (LPVs) are evolved red giant stars characterized by regular or semiregular luminosity variations driven by stellar pulsations. Their pulsation periods correlate strongly with both intrinsic brightness and stellar age, making them effective tools for distance estimation and for probing the structure of young to intermediate-age stellar populations in our and nearby galaxies.
The increasing availability of high-precision time-domain data from contemporary large-scale surveys enhances the observational importance of LPVs and calls for a more detailed empirical characterization of their variability and underlying physical parameters. At the same time, the abundance of such data provides crucial constraints for theoretical modeling by enabling systematic comparisons between theory and observations.

This review aims to provide a comprehensive overview of the current state of knowledge on LPVs, with particular emphasis on their observational characteristics, such as light curve morphology, period distributions, and variability amplitudes, as well as their applications in astrophysics, including distance determination and stellar population studies. The focus is placed on empirical findings derived from recent large-scale photometric surveys, rather than on the underlying theoretical models of stellar pulsation and evolution. For the latter, we refer the reader to e.g., \cite{Trabucchi2021,Ahmad2023} and references therein.

\section{MODERN DEFINITION AND CLASSIFICATION OF LONG PERIOD VARIABLES}

Long-period variables (LPVs) have yet to receive a universally accepted definition, so in this section, we explain how we interpret this term. Undoubtedly, the class of LPVs includes pulsating red giants near the tip of the asymptotic giant branch (AGB). Traditionally, this type of variable stars are classified into Miras, semi-regular variables (SRVs), and irregular variables \citep{Kholopov1985}. Miras are defined as variables with a total (peak-to-peak) amplitude of brightness variation in the {\it V}-band exceeding 2.5 mag, while SRVs exhibit smaller amplitudes. However, this criterion can be problematic for pulsators with amplitudes around 2.5 mag, as the changing shape of the light curves in LPVs can cause some stars to be classified as either Miras or SRVs depending on when they are observed. Another useful criterion for classifying LPVs is the number of radial modes they exhibit. Miras are typically single-mode variables pulsating in the fundamental mode, while SRVs most often exhibit two or three pulsation modes excited, usually the fundamental and first-overtone modes (see the sidebar titled STELLAR PULSATION).

\begin{marginnote}[]
\entry{Red giants}{evolved stars that have exhausted hydrogen in their cores, characterized by large radii and low surface temperatures}
\end{marginnote}

\begin{textbox}[b]\section{STELLAR PULSATION}
Stellar pulsation is a phenomenon observed across a wide range of stellar types, in particular in all red giants. The periodic expansions and contractions of star's outer layers provide insight into the internal structure of stars, making pulsations a central subject of asteroseismology. Pulsations can occur in different modes. When a star maintains spherical symmetry during oscillations, the pulsations are called {\bf radial}. In the {\bf fundamental radial mode}, the entire star expands and contracts as a whole. {\bf Overtone modes} occur when the oscillation pattern includes one or more radial nodes inside the star. When stellar pulsations break spherical symmetry, they are referred to as {\bf non-radial oscillations}.

Several mechanisms can drive stellar pulsations. The most important are the {\bf $kappa$} and {\bf $gamma$ mechanisms}, which are responsible for the variability of, among others, Cepheid and RR Lyrae stars. In these cases, changes in opacity within ionization zones trap and release
energy, sustaining the oscillations. Pulsations may also be excited stochastically by turbulent convection, leading to what are known as {\bf solar-like oscillations}.
\end{textbox}
\begin{marginnote}[]
\entry{Asymptotic giant branch}{a late evolutionary phase of low and intermediate mass stars after core helium exhaustion, marked by cooler surfaces, larger radii, and intensified mass-loss.}
\end{marginnote}

The existence of irregular variables as a subtype of LPVs has sparked controversy \citep{Percy2011}. Are there red giants in the upper part of the AGB that do not exhibit any periodicity? Our experience, drawn from the preparation of extensive catalogs of LPVs using photometric data from the Optical Gravitational Lensing Experiment (OGLE) survey, suggests that the light curves of all bright red giant stars display periodic variability to some extent. Red variables in which no periodicity is detected often turn out to be SRVs that have been observed for a too short timespan. Nevertheless, it is important to emphasize that an irregular component is present in the light curves of virtually all LPVs. SRVs that display persistent periodicity are classified as SRa stars, while those with poorly defined periods are classified as SRb stars.

Another class of stars commonly categorized as a subtype of LPVs are the pulsating red supergiants that exhibit semi-regular variability, known as SRc stars. In contrast, SRd stars are yellow giants and supergiants (of F, G, or K spectral type) that show semi-regular variations. However, since SRd stars are closely related to RV Tauri variables, which are classified as Type II Cepheids, we will not consider them as a subclass of LPVs in this work.

On the other hand, stars dubbed OGLE Small Amplitude Red Giants (OSARGs) by \cite{Wray2004} will be treated as a subclass of LPVs. OSARGs are red giant stars located on both the AGB and near the tip of the first-ascent red giant branch (RGB), with pulsation periods ranging from about 10 to over 100 days and amplitudes from millimagnitudes (the detection limit of ground-based survey telescopes) to a few tenths of a magnitude. Note that space-based telescopes, such as CoRoT and Kepler, have observed many fainter red giants (sometimes called $\xi$~Hydrae stars) with shorter periods and amplitudes below the millimagnitude level \citep[e.g.,][]{DeRidder2009,Bedding2010,Bedding2011,Yu2020}. However, in this paper, we restrict our attention to red giants whose variability can be studied from the ground, i.e. OSARGs, SRVs, and Miras.

\begin{figure}[t]
\includegraphics[width=6in]{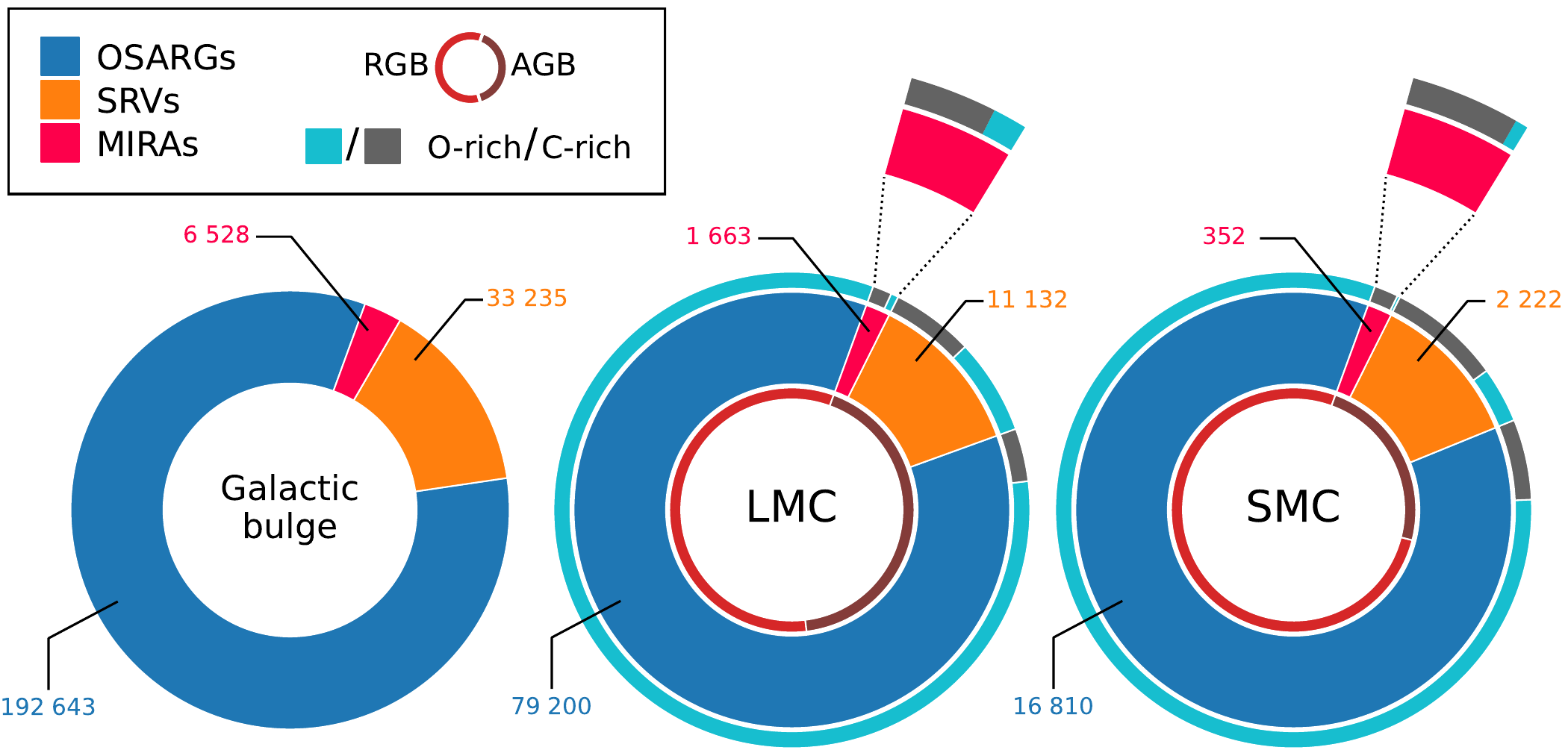}
\caption{The modern classification of LPVs into Miras (red), SRVs (orange), and OSARGs (blue), showing proportions among these LPV subclasses. Object counts are based on OGLE-III catalogs of LPVs in the Galactic bulge \citep{Soszynski2013}, the Large Magellanic Cloud \citep[LMC;][]{Soszynski2009}, and the Small Magellanic Cloud \citep[SMC;][]{Soszynski2011}. For the LMC and SMC, the outer circle shows a fraction of O-rich (cyan) and C-rich (grey) stars, while the inner circle indicates a fraction of RGB (dark red) and AGB (brown) stars, within each class. Figure courtesy of Jan Skowron.}
\label{fig:classification}
\end{figure}

In summary, the class of LPVs includes pulsating red giant and supergiant stars with a wide range of masses (from a fraction to several dozen solar masses), evolving on the upper RGB or AGB, with periods ranging from days to years and amplitudes of light variation from millimagnitudes to several magnitudes.
The classification scheme adopted throughout this review is schematically presented in \textbf{Figure~\ref{fig:classification}}, while
\textbf{Table~\ref{tab:classes}} provides a short summary of the defining characteristics, typical properties, and distinguishing features of each class of LPVs. This table may serve as a convenient reference for the reader to quickly compare the observational and physical attributes of Miras, SRVs, and OSARGs discussed throughout the text.

\begin{table}[bht]
\tabcolsep10pt
\caption{Characteristic properties of LPV subclasses}
\label{tab1}
\begin{center}
\begin{tabular}{@{}l|c|c|c|c|c@{}}
\hline
\vspace{-0.04cm}
\textbf{LPV} & \textbf{pulsation} & \textbf{pulsation} & \textbf{pulsation} & \textbf{evolutionary} & \textbf{chemistry} \\
& \textbf{periods} & \textbf{amplitudes}$^{\rm a}$ & \textbf{modes} & \textbf{stage} & \\
\hline
\textbf{Mira} & $\gtrsim 80$ days & $A_V >2.5$ mag & single-mode & AGB & O/C-rich\\
              &                   & $A_I >0.8$ mag &   radial    &     &         \\
              &                   &                & fundamental &     &         \\
\hline
\textbf{SRV}  & $\gtrsim 20$ days & $A_V <2.5$ mag &  multi-mode & AGB & O/C-rich\\
              &                   & $A_I <0.8$ mag &    radial   &     &         \\
              &                   &                & fundamental/overtone &   &   \\
\hline
\textbf{OSARG} & $\sim10 - 100$ days & $A_I \lesssim 0.2$ mag &   multi-mode  & RGB/AGB & mainly\\
               &                     &                        &  radial/non-radial &    & O-rich\\
\hline
\end{tabular}
\end{center}
\begin{tabnote}
$^{\rm a}$LPVs often exhibit irregular fluctuations that exceed their pulsation amplitudes.
\end{tabnote}
\label{tab:classes}
\end{table}

\section{PHYSICAL PROPERTIES OF PULSATING RED GIANT STARS}

\subsection{Evolutionary stage}

Red giant stars which have exhausted helium in their core start moving along the AGB in the Hertzsprung-Russell diagram. In the course of the subsequent evolution the core contracts and becomes hotter, the star's photosphere expands and its luminosity increases, while the star becomes cooler. At this stage the helium burning in the core is completed and the nuclear energy comes from burning helium in a shell surrounding a carbon and oxygen core and from hydrogen fusion in another outer shell. This evolutionary path is undertaken by low- and intermediate-mass stars, from about 0.5 to 8 M$_\odot$ \citep{Hofner2018}.

\begin{marginnote}[]
\entry{Hertzsprung–Russell diagram}{a plot of stellar luminosity versus surface temperature}
\end{marginnote}

The AGB phase consists of two stages. In the early AGB (E-AGB), the star is powered by helium fusion in a shell surrounding a degenerate carbon–oxygen core. As helium is exhausted, the star enters the thermally pulsing AGB (TP-AGB) phase, during which hydrogen shell burning dominates. Over time, helium produced by hydrogen burning accumulates until it triggers periodic and short-lived thermal instabilities, known as thermal pulses or helium shell flashes. These episodes, which occur on timescales of 10,000 to 100,000 years, cause a temporary increase in luminosity and structural expansion, followed by deep convective mixing, called the third {\it dredge-up}, which transports nucleosynthetic products, such as s-process elements, to the surface. Repeated dredge-ups can progressively enrich the stellar atmosphere, sometimes leading to the development of carbon-rich AGB stars.

Miras and SRVs can be classified as oxygen-rich (O-rich; C/O $<$ 1) or carbon-rich (C-rich; C/O $>$ 1), depending on the relative abundance of oxygen to carbon in their surface composition. Evolutionary models predict that stars entering the AGB are O-rich; however, third dredge-up episodes gradually enrich the atmospheres of giants with carbon, ultimately transforming them into C-rich stars. This transformation radically changes the spectrum of the star, but it also affects the shapes of the light curves of Miras and SRVs and shifts their positions on some color--color and period--luminosity diagrams.
The balance between O-rich and C-rich stars is also metallicity-dependent: in higher metallicity environments O-rich Miras and SRVs dominate, whereas in lower-metallicity galaxies C-rich stars are far more prevalent (see the comparison between the Magellanic Clouds in \textbf{Figure~\ref{fig:classification}}).

As already noted, practically all luminous AGB stars undergo pulsations, which are a natural consequence of the stars extended, low-gravity envelopes. These pulsations become more pronounced as the star evolves along the AGB -- initially pulsating in overtone modes with low amplitudes and shorter periods, and eventually high-amplitude fundamental-mode pulsations characteristic of Mira variables near the tip of the AGB. The Mira phase corresponds to the peak in both stellar luminosity and mass-loss rate.
The pulsational variability is also observed in red supergiants, which are more massive than typical LPVs (with masses up to $\sim25$ M$_\odot$), being in the late core helium-burning phase of their evolution.

\subsection{Stellar winds}

AGB stars have very expanded atmospheres, with typical radii of several hundred R$_\odot$, which combined with masses below 8 M$_\odot$, result in low surface gravities that cause these stars to lose mass in the form of a slow (5–30 km/s) stellar wind. Shock waves resulting from convection and stellar pulsations can enhance the mass-loss rate, which is between $10^{-8}$ to $10^{-5}$ M$_\odot$/year, and can reach as much as $10^{-4}$ M$_\odot$/year \citep{Hofner2018}, making them the major source of dust in the Milky Way \citep{Tielens2005}.
Numerous studies of the past decades have addressed the nature of the AGB winds, such as the mass-loss rate, velocity, chemical composition, and dust-to-gas ratios, 
thoroughly reviewed by \cite{Decin2021}. 
Recent advances in three-dimensional hydrodynamical modelling further enable more detailed studies of AGB stars and their dust-driven winds with a variety of modern codes \citep[e.g.,][and references therein]{Siess2022,Esseldeurs2023,Freytag2023}.

In binary systems, in which one of the stars is in the AGB phase, the mass transfer takes place due to the strong winds, either via Roche-lobe overflow (in close binaries) or more isotropically (in wide binaries), and leads to the formation of various types of interesting objects, such as blue stragglers \citep{Mathieu2015}, symbiotic binaries \citep{deValBorro2009} or carbon enhanced metal poor stars \citep{Beers2005}. It is therefore not surprising that numerous studies have tried to understand the process of mass transfer and the interaction of the wind with the companion in the binary system 
\citep[e.g.,][]{Theuns1993,Mohamed2007,Chen2017,Liu2017,Saladino2019,Maes2021,Malfait2021,Aydi2022}. 
One of the possible effects of such interaction are dusty spiral structures following the companion on its orbit, associated with the matter lost by the AGB star via stellar winds. Indeed, similar structures have been observed at millimeter wavelengths with ALMA (the ATOMIUM programme), and their presence has been attributed to binary interactions \citep{Decin2020}. A recent analysis of the ATOMIUM data for 14 AGB and 3 RSG stars suggests the presence of the companion for 11 of the 14 AGB stars, while an increased mass-loss rate around one target is explained by additional dust formed around the host due to shocks generated by the passage of its close companion \citep{Danilovich2025}. Such companion-enhanced dust formation mechanism (in addition to the pulsation-enhanced dust formation) would explain a variety od dust features observed around AGB stars.

The mass transfer onto the companion can also result in the substantial increase of its mass \citep{Chen2017,Saladino2019}, and e.g., in the case of planets orbiting red giants, they indeed tend to be more massive than planets around main sequence stars \citep{Jones2014,Niedzielski2015}. The possibility of transforming Jupiter-like planets into brown dwarfs is one of the considered scenarios that takes place during the AGB evolution of the host star of a planetary system \citep{Livio1984,Retter2005}. However, this scenario has not been further addressed in the recent years and there are currently no detailed models of AGB star-planet mass transfer.
\begin{marginnote}[]
\entry{brown dwarfs}{substellar objects with masses of $\sim 13-80$ Jupiters ($0.012-0.075 \; {\rm M}_\odot$), insufficient to sustain stable hydrogen fusion}
\end{marginnote}

\section{LIGHT-CURVES OF LONG PERIOD VARIABLES}

The main criteria for classifying a variable star are its light-curve morphology, period, and amplitude. For LPVs, long-term monitoring over many years is particularly important, as cycle-to-cycle variations are often a key classification feature.
The observed light-curve morphology also depends on the photometric filter: classification is most straightforward in the optical bands, where amplitudes are larger and shapes more distinct. At longer wavelengths, light curves become sinusoidal, making it more difficult to distinguish between different classes of variable stars.
In this section, we highlight the rich diversity of LPV light curves observed over past decades by the OGLE survey.

\subsection{Miras}

\begin{figure}[h]
\includegraphics[width=5.6in]{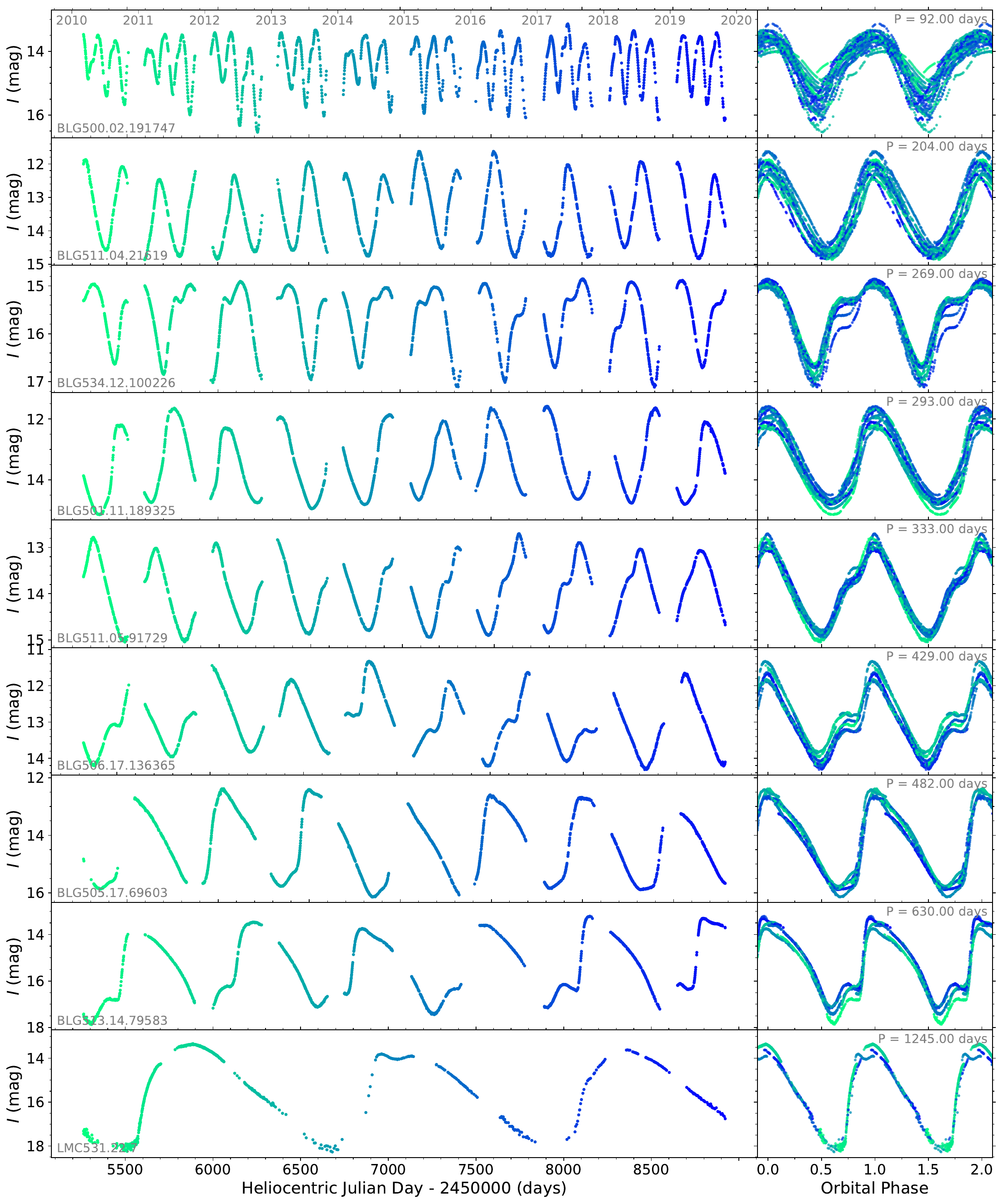}
\caption{I-band light curves of selected O-rich Miras obtained between 2010 and 2020 during OGLE-IV survey. Left panels show unfolded time-series photometry, while right panels show the same data folded with the pulsation periods. The stars are ordered by pulsation period (given in the upper right corner).}
\label{fig:o-miras}
\end{figure}

\begin{figure}[h]
\includegraphics[width=5.6in]{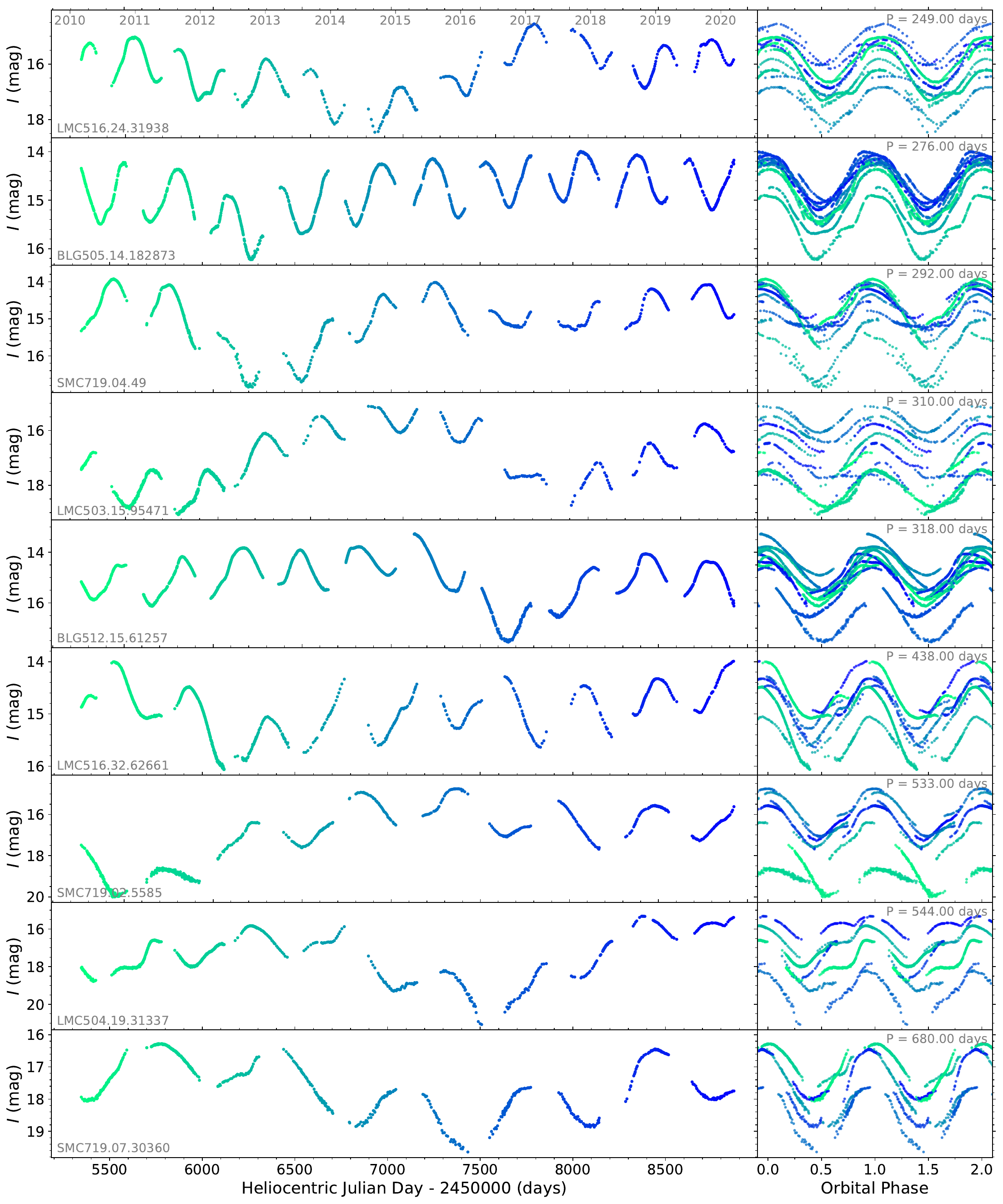}
\caption{I-band light curves of selected C-rich Miras obtained between 2010 and 2020 during OGLE-IV survey. Left panels show unfolded time-series photometry, while right panels show the same data folded with the pulsation periods. The stars are ordered by pulsation period (given in the upper right corner).}
\label{fig:c-miras}
\end{figure}

Mira variables are among the most prominent LPVs due to their large brightness variations, particularly at optical wavelengths. Their pulsation periods range from roughly 80 to over 1000 days, with V-band amplitudes reaching from 2.5 up to 8–9 magnitudes and I-band amplitudes from 0.8 up to 4–5 magnitudes. 
\textbf{Figures~\ref{fig:o-miras}} and~\textbf{\ref{fig:c-miras}} show I-band light curves of several Mira stars observed over a 10-year period. The light curves of the O-rich Miras (\textbf{Figure~\ref{fig:o-miras}}) are generally stable over time, but the light curves of the C-rich Miras (\textbf{Figure~\ref{fig:c-miras}}) often display slow, irregular changes in mean brightness, likely caused by obscuration from circumstellar dust shells.

Many Miras exhibit noticeable humps or bumps along the rising part of their light curves. These features are thought to result from a 2:1 resonance between the fundamental and first-overtone pulsation modes. In some cases, the humps are pronounced enough that the light curve appears to display two maxima. Furthermore, the prominence of these features often varies from one pulsation cycle to the next.

\subsection{Semi-regular variables}

\begin{figure}[h]
\includegraphics[width=5.7in]{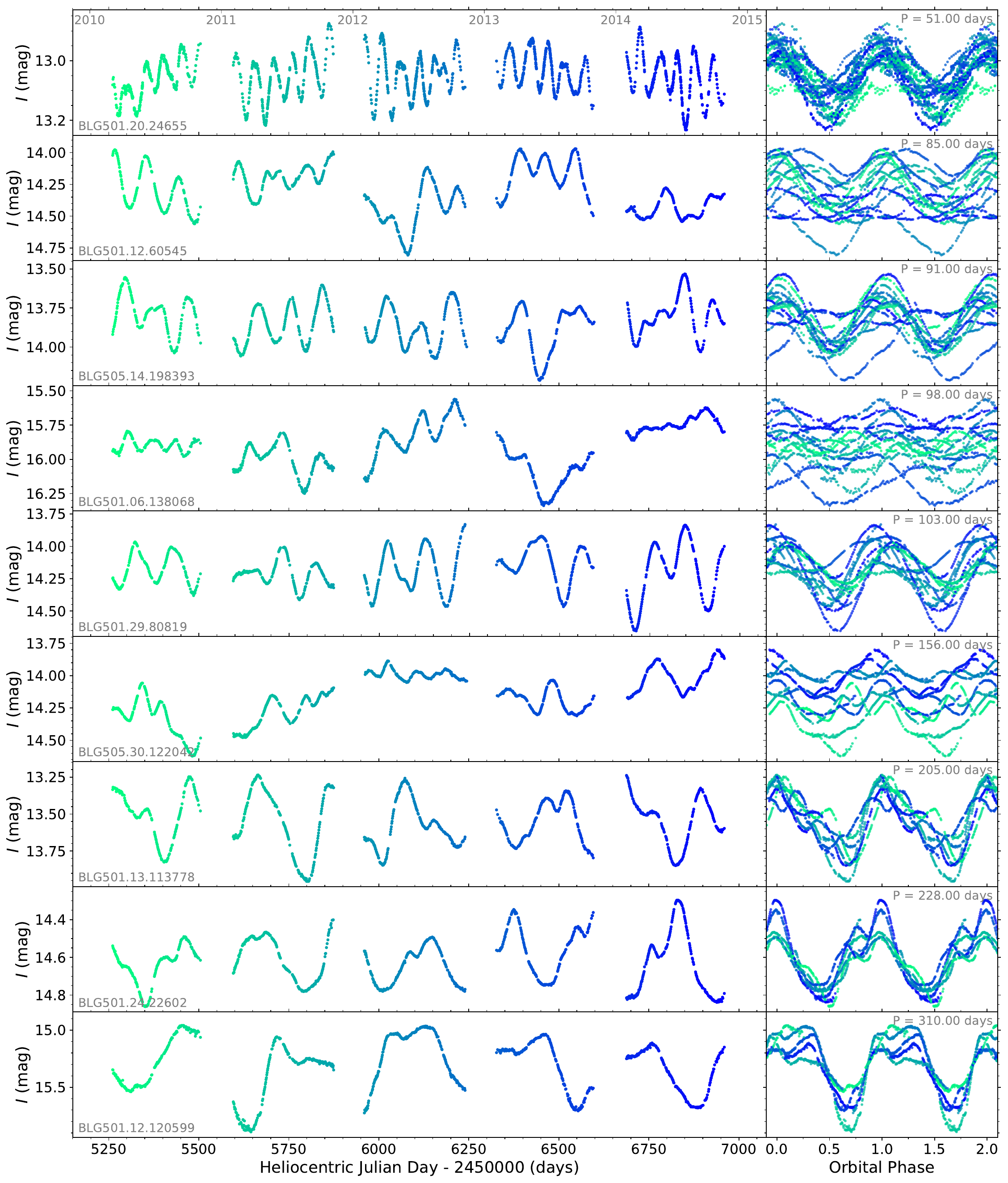}
\caption{Typical light curves of SRVs. Left panels show unfolded light curves collected during the first 5 years of the OGLE-IV survey (2010-2015). Right panels show the same light curves folded with the primary (the largest-amplitude) period. The stars are ordered by pulsation period (given in the upper right corner).}
\label{fig:srvs}
\end{figure}

SRVs have pulsation periods that span approximately 20 to 500 days and display a combination of regular cycles and irregular variations, as shown in \textbf{Figure~\ref{fig:srvs}}. Even their periodic behavior can show large changes in both phase and amplitude, and some stars exhibit additional long secondary periods. Traditionally, SRa stars are distinguished from Miras by having visual amplitudes below 2.5 magnitudes in the V-band (0.8 magnitudes in the I-band), although this is an artificial boundary and individual stars may shift between classifications depending on the observational time span. The main distinction from Miras lies in the pulsation modes: most SRVs simultaneously oscillate in both the fundamental and first-overtone modes (see Section~\ref{sec:pl}).

\subsection{OGLE small-amplitude red giants}

OSARGs exhibit pulsation periods ranging from roughly 10 to just above 100 days. Their amplitudes are closely linked to absolute brightness, with the faintest stars showing variations of only a few millimagnitudes, just above the detection threshold of OGLE ground-based photometry, while the brightest OSARGs can display periodic variations of around 0.2 magnitudes in the I-band, often accompanied by additional irregular fluctuations of larger amplitude. Furthermore, at least 30\% of OSARGs exhibit long secondary periods.

\begin{figure}[h]
\includegraphics[width=5.6in]{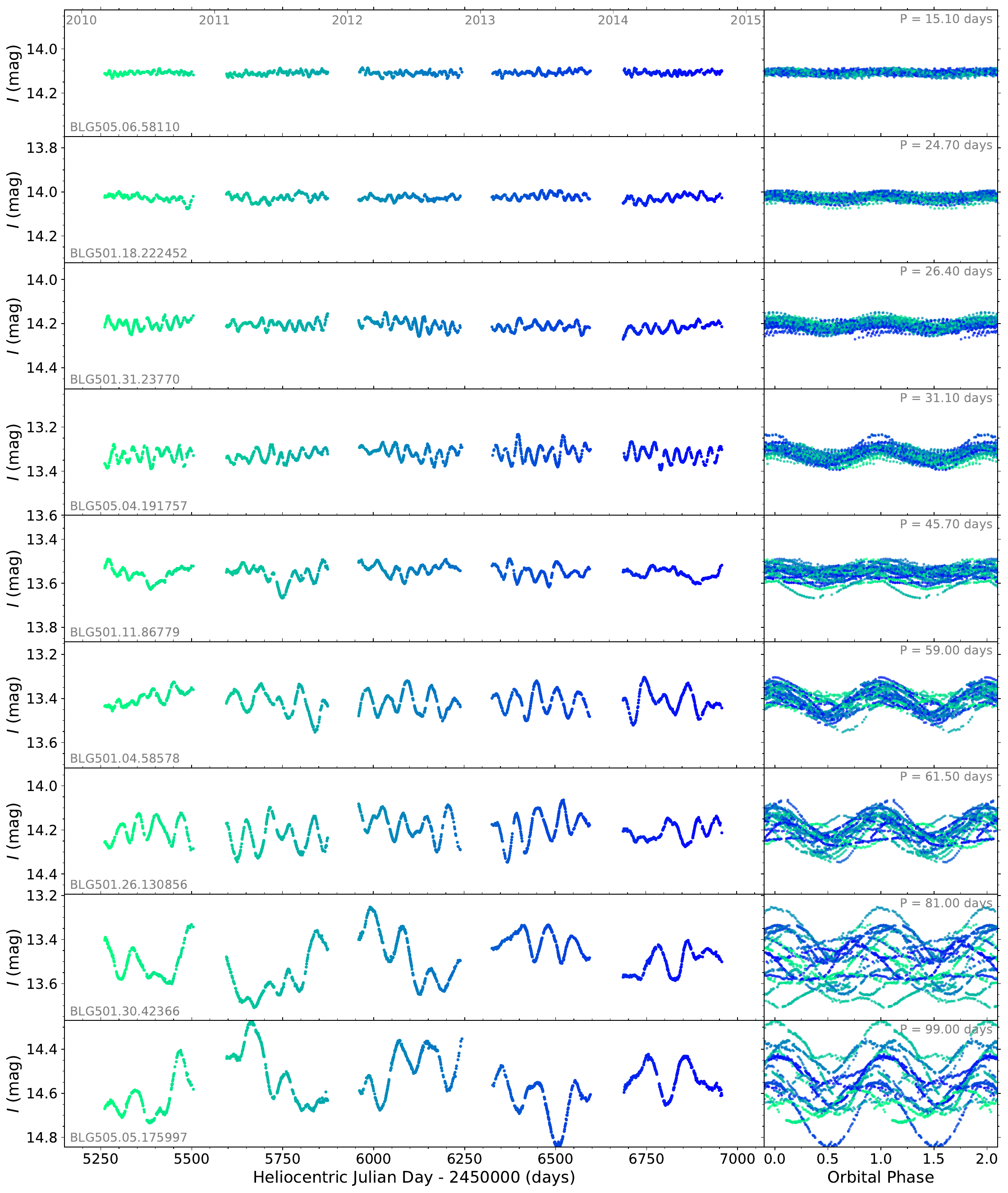}
\caption{A sample of OSARG light curves obtained during the first five years of the OGLE-IV project (2010–2015). The left panels display unfolded time-series, while the right panels show the same data folded with the primary pulsation period. The stars are ordered by pulsation period (given in the upper right corner). Note that the magnitude scale is the same for all stars, to emphasize the variety of OSARG amplitudes.}
\label{fig:osargs}
\end{figure}

Examples of OSARG light curves are shown in \textbf{Figure~\ref{fig:osargs}} (to illustrate the variety of amplitudes, each light curve is plotted with the same magnitude range). They are multiperiodic and generally resemble those of SRVs. This is expected, as OSARGs evolve into SRVs as stars ascend the AGB, and later develop into Miras. Despite this, OSARGs and SRVs follow different period–luminosity relations (see Section~\ref{sec:pl}) and typically exhibit a larger number of excited pulsation modes, often oscillating simultaneously in three, four, or more radial and non-radial modes.

\subsection{Long secondary period variables}

The long secondary period (LSP) is observed in at least one-third of LPVs. Its duration spans several months to a few years, typically about ten times longer than the dominant pulsation periods of the same stars. At visual wavelengths, the modulation can reach amplitudes of about one magnitude, often exceeding those of the primary pulsations.

\begin{figure}[h]
\includegraphics[width=5.7in]{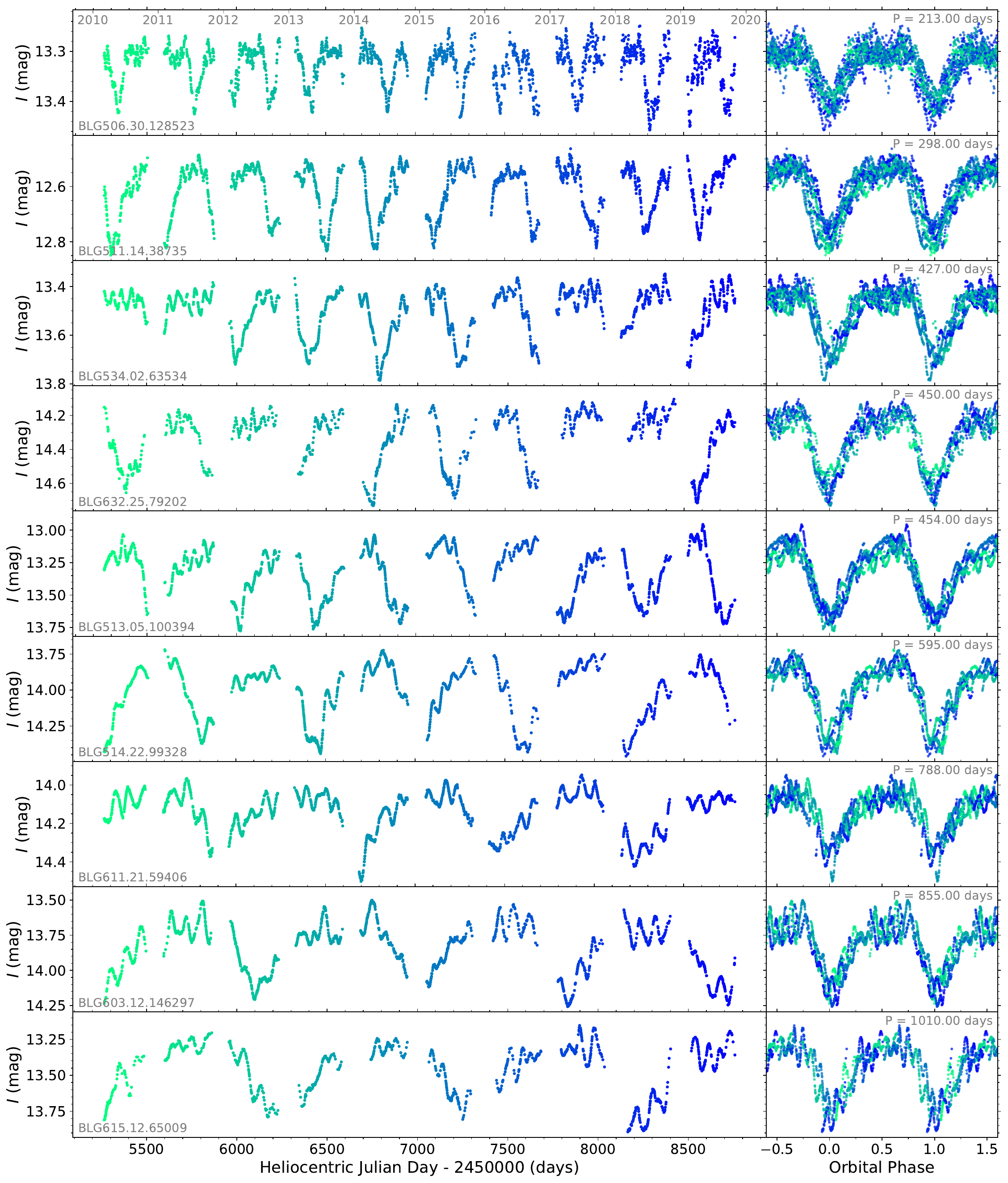}
\caption{A few examples of the LSP light curves obtained between 2010 and 2019 during OGLE-IV survey. The left panels display unfolded time-series, while the right panels show the same data folded with the long secondary period. The stars are ordered by long secondary period (given in the upper right corner).}
\label{fig:lsps}
\end{figure}

Examples of LSP light curves with relatively stable amplitudes of the LSP variability and small amplitudes of stellar pulsations (visible as short-period oscillations in the light curves) are shown in \textbf{Figure~\ref{fig:lsps}}. The characteristic variability pattern consists of two phases: one in which the brightness remains nearly constant (aside from pulsations) or increases slowly with time, and another characterised by a triangular eclipse. Each phase covers roughly half of the LSP cycle, but in the largest-amplitude LSP variables, the duration of the triangular minimum may extend beyond half the period, while the maxima often take a rounded shape. In some LSP stars, the amplitudes of the light curves change noticeably from cycle to cycle, and the LSP variability may even temporarily vanish.

\section{OBSERVATIONAL DATA}
\label{sec:data}

A thorough understanding of the properties of LPVs requires time-series observations spanning years or preferably decades, not only because of the long periods but also due to the random cycle-to-cycle fluctuations present in the light curves of red giant and supergiant stars. The history of LPV observations dates back to 1596, when David Fabricius (1564-1617) discovered that Omicron Ceti (Mira) varies in brightness. However, until the second half of the 19th century, discoveries of variable stars remained rare and largely accidental. In the 200 years following the discovery of Mira Ceti, only four additional LPVs were identified, including three Miras and one SRV \citep{Hoffleit1997}. Only with the advent of photographic techniques did the number of known LPVs rise to several hundred by the end of the 19th century, followed by the discovery of thousands more in the 20th century.

An invaluable collection of long-term visual light curves is available in the American Association of Variable Star Observers (AAVSO) International Database, which contains observations gathered by thousands of amateur astronomers over more than a century. The AAVSO data have been used to study secular period changes in Miras \citep{Templeton2005,Merchan2023} and to detect long secondary periods, reaching several thousand days, in red supergiants \citep{Kiss2006}. This type of research requires continuous observations spanning at least several decades.

About 30 years ago, a breakthrough occurred in the study of variable stars, including LPVs. Large-scale time-domain sky surveys were introduced, with the aim of monitoring millions and later even billions of stars, rather than focusing on individual objects as was done before. Since then, it has become possible to precisely investigate the statistical properties of LPVs, including their distributions in the period--luminosity, color--luminosity, and period--amplitude planes. 

The pioneering time-domain optical sky surveys were originally devoted to the search for gravitational microlensing events, but their extensive photometric databases proved indispensable for discovering and studying variable stars (Paczyński 1986). Observations from the MACHO microlensing survey were used to identify a series of period--luminosity relations exhibited by LPVs \citep{Cook1997,Wood1999,Wood2000}, while data from the EROS \citep{Cioni2001}, OGLE \citep{Kiss2003,Kiss2004,Ita2004,Soszynski2004a,Soszynski2004b}, and MOA \citep{Noda2004} projects helped to refine this distribution. At present, several extensive catalogs of LPVs are available to the astronomical community, including those published by ASAS \citep[e.g.][]{Pojmanski2002}, OGLE \citep[][]{Soszynski2009,Soszynski2011,Soszynski2013,Iwanek2023}, ASAS-SN \citep[e.g.][]{Jayasinghe2018}, ATLAS \citep[e.g.][]{Heinze2018}, ZTF \citep[e.g.][]{Chen2020b}, CATALINA \citep{Drake2017}, LAMOST \citep{Qiao2024} and the near-infrared VVV \citep[e.g.][]{Nikzat2022,Albarracin2025}, and Palomar Gattini-IR \citep{Suresh2024} surveys, although not all of them provide a detailed classification into subclasses of LPVs.
In the majority of these catalogs, stellar classification is performed automatically using a variety of computational methods. By contrast, only the ASAS and OGLE catalogs implement manual verification as the final step of the classification process, whereby each light curve is carefully inspected by a human, yielding the highest levels of purity and completeness among these datasets. Consequently, the OGLE catalogs of LPVs, containing over 111,000 stars in the Magellanic Clouds \citep{Soszynski2009,Soszynski2011} and over 298,000 in the Milky Way bulge and disk \citep{Soszynski2013,Iwanek2022} are considered a gold standard for a wide range of astrophysical studies.

The largest catalog to date was published by \citet{Lebzelter2023} as part of Gaia Data Release 3 (DR3). It includes more than 1.7 million automatically selected LPV candidates, with periods determined for about 390,000 of them. The Gaia catalog was later supplemented with radial-velocity time series for 9,614 LPVs \citep{Gaia2023}, representing the largest such dataset published to date. These valuable data offer a preview of what will be included in future releases of the Gaia mission.

Also the wealth of spectroscopic data for LPVs from large-scale surveys such as APOGEE \citep{Blanton2017}, GALAH \citep{DeSilva2015}, LAMOST \citep{Cui2012}, and RAVE \citep{Casey2017} provides an opportunity to investigate their chemical properties and thereby gain insight into evolutionary processes, such as e.g., the third dredge-up episodes \citep[e.g.,][]{Jayasinghe2021}. However, this extensive spectroscopic resource is far from being fully exploited.

\section{PERIOD--LUMINOSITY AND PERIOD--AGE RELATIONS}

\subsection{Period--Luminosity Relations}
\label{sec:pl}

\citet{Gerasimovic1928} was the first to discover a relationship between the periods of Mira variables and their visual magnitudes, revealing that shorter-period Miras are on average brighter than those with longer periods. \citet{Wilson1942} confirmed this dependence but noticed its large dispersion. Indeed, the correlation between the periods and visual absolute magnitudes of LPVs is rather weak, so it cannot be considered a strict period--luminosity relation, as observed, for example, for classical Cepheids.

\citet{Eggen1975} suggested that Miras follow a much tighter period--luminosity relationship when bolometric magnitudes are used. However, accurate determination of the bolometric magnitudes requires observations in the infrared domain, near the peak in the energy distribution of cool giant stars. The progress in infrared astronomy that took place in the 1970s made it possible to show that Miras indeed follow a narrow period--luminosity relation in the infrared bands and, consequently, in bolometric luminosities as well. \citet{Glass1981} were the first to demonstrate a tight period--bolometric magnitude relation based on 12 Mira variables in the LMC. Since then, most of the significant discoveries related to the distribution of LPVs on the period--luminosity diagram have been driven by studies of the Magellanic System. \citet{Feast1989} determined near-infrared period--luminosity relation for Mira stars based on 29 O-rich and 20 C-rich Miras in the LMC, finding that both spectral types follow very similar relations in the {\it K}-band. They also noticed that O-rich Miras with periods exceeding 420 days are brighter than expected from the extrapolation of the period--luminosity relation fitted to shorter-period Miras. Years later, this observation was explained by the phenomenon of "hot-bottom burning", which occurs in relatively massive AGB stars \citep{Whitelock2003}.

\citet{Wood1996} were the first to notice that some SRVs follow a different period--luminosity relation than Miras. The newly discovered period--luminosity sequence was parallel to the one populated by Miras but corresponded to periods shorter by a factor of 2.2. This provided a strong argument that Miras pulsate in the fundamental mode, while SRVs usually have both fundamental and first-overtone modes excited.

A true revolution in the period--luminosity diagram for long-period variables came with the advent of large-scale sky surveys such as MACHO \citep{Alcock1992} and OGLE \citep{Udalski1992}, whose original goal was to detect the first gravitational microlensing events. Particularly fruitful were the long-term observations of thousands of red giant stars in the LMC, which revealed that LPVs follow multiple period-–luminosity relations \citep{Cook1997}. \citet{Wood1999} and \citet{Wood2000} distinguished five period--luminosity sequences and labeled them with the letters A through E.

The research carried out in the following years \citep{Kiss2003,Kiss2004,Ita2004,Soszynski2004a,Soszynski2004b}, mainly based on photometric data from the MACHO and OGLE projects, expanded the number of period--luminosity relations followed by LPVs. Eventually, \cite{Soszynski2007} identified fourteen such relations, with separate sequences formed by O-rich and C-rich Miras and SRVs, as well as RGB and AGB OSARGs. Aside from the OGLE and MACHO surveys, photometric data from the EROS-2 \citep{Spano2011}, Gaia \citep{Lebzelter2023,Chen2024}, and ATLAS \citep{Hey2025} projects have also been used to study the distribution of LPVs on the period--luminosity diagram. Recent developments in the theory of red giant pulsation \citep{Trabucchi2021,Ahmad2025} enable an increasingly accurate reproduction of the pattern observed on the period--luminosity plane.

\begin{figure}[p]
\includegraphics[width=5.4in]{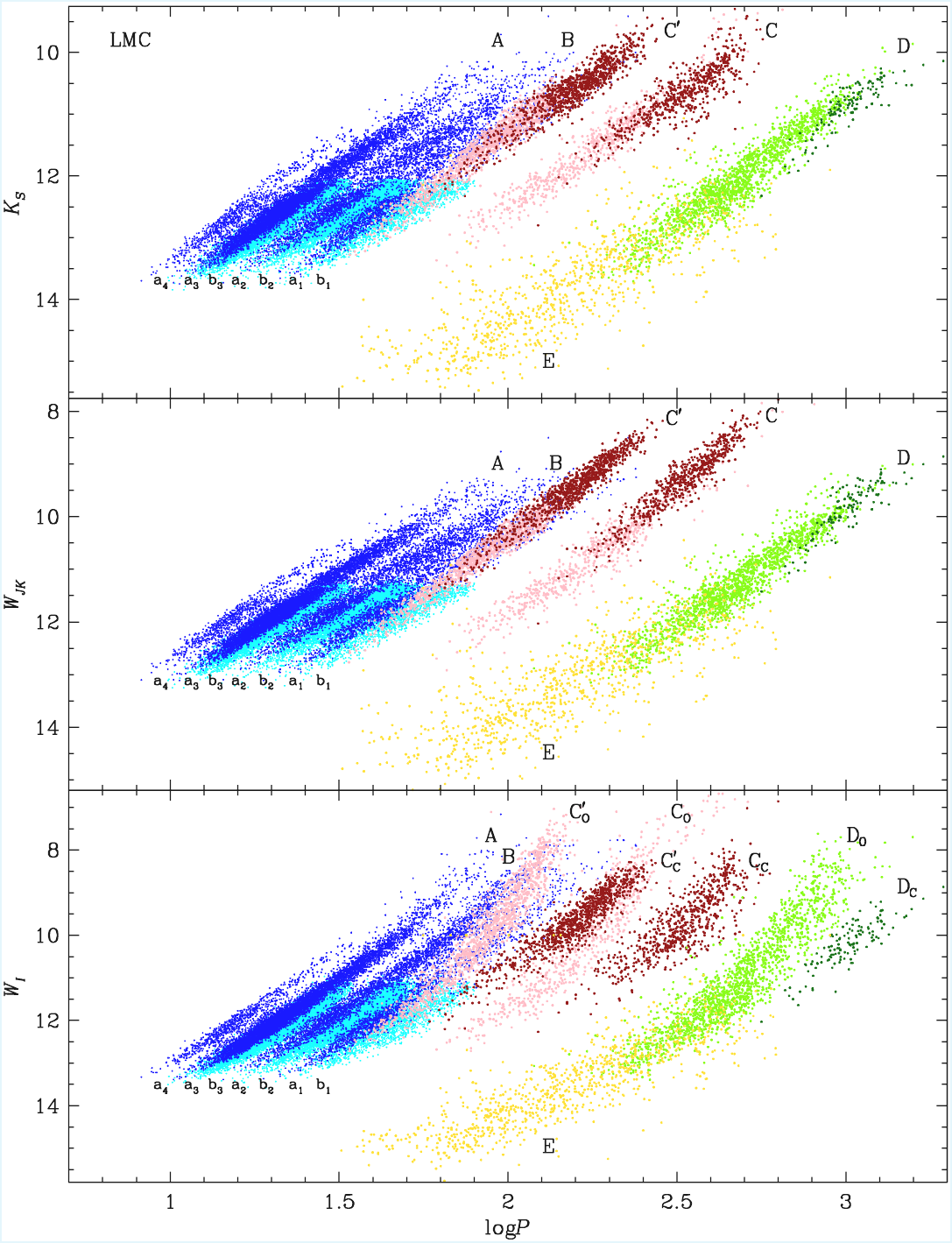}
\caption{Period--luminosity and period--Wesenheit diagrams of LPVs in the LMC \citep{Soszynski2007}. OSARG variables, occupying sequences A and B, are shown as blue points (RGB as light blue, AGB as dark blue). Miras (sequence C) and SRVs (sequences C and C') are marked with pink (O-rich) and red (C-rich) points. Light and dark green points refer to O-rich and C-rich LSP variables (sequence D), respectively. Yellow points indicate ellipsoidal red giants (sequence E).}
\label{fig:PL}
\end{figure}

\textbf{Figure~\ref{fig:PL}} shows near-infrared period--luminosity and period--Wesenheit diagrams for LPVs in the LMC. The periods of the stars were calculated based on observations conducted as part of the OGLE project. The infrared Wesenheit index is defined as $W_{JK}=K_\mathrm{s}-0.686(J-K_\mathrm{s})$, while optical Wesenheit index is defined as $W_{I}=I-1.55(V-I)$. The $J$- and $K_\mathrm{s}$-band magnitudes originate from the IRSF Catalog \citep{Kato2007} or the 2MASS Catalog \citep{Cutri2003}, while $I$- and $V$-band magnitudes originate from the OGLE project.

\begin{marginnote}[]
\entry{Wesenheit index}{a reddening-free combination of magnitudes and colors, designed to correct for interstellar extinction based on an assumed total-to-selective extinction ratio.}
\end{marginnote}

According to the nomenclature introduced by \citet{Wood1999} and other authors, Miras lie on period--luminosity sequence C, while semiregular variables occupy sequences C and C$'$, corresponding to the fundamental and first-overtone pulsation modes, respectively \citep{Wood1996}. O-rich and C-rich variables follow similar period--luminosity relations in the near-infrared $K_\mathrm{s}$ band (upper panel of Figure~\ref{fig:PL}), although some C-rich Miras lie significantly below sequence C, likely due to strong attenuation by circumstellar dust \citep{Yuan2017}. Indeed, when the $K_\mathrm{s}$ band is replaced by the extinction-free near-infrared Wesenheit index (middle panel of Figure~\ref{fig:PL}), these outlying C-rich Miras align along a much narrower ridge. In turn, in the optical period--Wesenheit diagram (lower panel of Figure~\ref{fig:PL}), O-rich and C-rich LPVs separate into roughly parallel sequences. This feature can be used to identify the two chemical types of AGB stars.

OSARG variables follow a distinct set of period--luminosity sequences compared to SRVs. At least four such sequences can be identified for OSARGs on the AGB \citep{Soszynski2007}, while OSARGs on the RGB form three sequences that are slightly offset relative to their AGB siblings \citep{Kiss2003}. OSARGs are multi-periodic variables; however, their dominant (largest-amplitude) periods most commonly fall on sequences B and A, which correspond to radial pulsations in the first and second overtones \citep{Mosser2013,Yu2020}. In addition, low-order non-radial modes appear as closely spaced additional periods, which split sequences A and B into three parallel period--luminosity relations \citep{Soszynski2007,Wood2015}. According to the scenario proposed by \citet{Wood2015}, AGB stars evolve through successive period--luminosity sequences (A, B, C$'$, and C), while simultaneously increasing their luminosity.

Note that sequences B and C$'$ are shifted relative to each other, although they correspond to pulsations of red giant stars in the same first-overtone mode, suggesting a fundamental property that distinguishes OSARGs from SRVs. \citet{Trabucchi2025} proposed that the difference between these two types of LPVs lies in the mechanism responsible for exciting their pulsations. The oscillations of OSARGs are thought to be driven by a solar-like stochastic convective process, while the pulsations of SRVs and Mira variables are believed to be powered by the ${\kappa}$-mechanism, although with a significant and not fully understood role of convection \citep{Ahmad2025}.

The broad Sequence E, located in the lower part of the period--luminosity diagrams shown in \textbf{Figure~\ref{fig:PL}}, is formed by close binary systems containing red giant stars. These are eclipsing and ellipsoidal variables in which the red giant nearly fills its Roche lobe, which results in brightness variations caused by the rotation of the star distorted by the tidal forces of its companion. The greater the amplitude of the ellipsoidal variations, the narrower the width of Sequence E -- reflecting the extent to which the Roche lobe is filled.

About 10–20\% of ellipsoidal red giants are found in binary systems with eccentric orbits \citep{Soszynski2004b}. Variables of this type are referred to as heartbeat stars \citep{Thompson2012}, as their light curves resemble electrocardiogram signatures. The higher the orbital eccentricity, the longer the orbital period at a given luminosity (and thus size) of the red giant \citep{Wrona2022}. Consequently, heartbeat stars typically appear to the right of sequence E on the period--luminosity diagram.

Finally, Sequence D is associated with the most mysterious phenomenon observed in red giant stars in the upper RGB and AGB -- the so-called Long Secondary Periods (LSPs). This phenomenon remains without a definitive explanation, although the fact that Sequence D extends from Sequence E toward higher luminosities and longer periods \citep{Soszynski2004a,Soszynski2004b} suggests a connection between LSPs and binarity. Further discussion of LSPs in red giant stars can be found in Section~\ref{sec:lsp} of this work.

\subsection{Period--Age Relations}

Observational evidence acquired over the last century has shown that LPVs follow a period–age relation, where younger stars pulsate with longer periods than their older counterparts. This correlation was first hinted at through early kinematic studies in the solar neighborhood, which found that LPVs with shorter periods move differently than longer-period ones. In particular, they display greater velocity dispersion, increased lag relative to the local standard of rest, and a higher vertical distribution above the Galactic plane \citep{Feast1963}. These properties indicate that shorter-period LPVs belong to older stellar populations. This idea has since been reinforced by studies of LPVs in various Galactic environments, including globular clusters, the Galactic bulge, and nearby dwarf galaxies. The observed relation is also connected to stellar metallicity, further linking pulsation characteristics to evolutionary status. 
Later investigations, supported by a Hipparcos space-based mission, have strengthened the empirical foundation of the period--age relation \citep{Feast2007}, and a plethora of data from large-scale surveys enabled its use in detailed studies of LPV motions and spatial distribution, which will be discussed in the following chapter.

Even though the theoretical aspect of LPVs is not the scope of this review, it is worth mentioning here that, despite strong observational support, the theoretical modeling of the period--age relation in LPVs has received little attention in the past decades \citep{Eggen1998,Wyatt1983}, even though it was predicted quite early by \cite{Kippenhahn1969} for classical Cepheids. However, recent advancements in the modeling of evolved stars have resulted in solid theoretical grounds supporting the period--age relation \citep{Trabucchi2022}. 
Regrettably, the period--age relation for LPVs shows significant spread due to structural changes during the short but complex AGB phase, especially stellar expansion which affects the pulsation periods. This spread limits the precision of LPVs as age indicators unless carefully modeled. As a result, studies often focus only on Miras, whose brief and well-defined evolutionary phase narrows the period range at a given age. However, this excludes many SRVs that pulsate in the fundamental mode and could also serve as useful age tracers.

The availability of extensive LPV catalogs has motivated numerous studies investigating the reliability and applicability of the period--age relation.
Traditionally, Mira ages have been inferred by associating them with disc stellar populations of comparable velocity dispersions \citep[e.g.,][]{Feast2006}. In their analysis of bulge properties, \cite{Lopez2017} provided a quadratic fit of the period--age relation to the available literature data for Galactic Miras. 
\cite{Nikzat2022} later upgraded the relation of \cite{Lopez2017}, which showed a better agreement with the available data, to compensate for its unphysical behaviour at the shortest and longest period.
In a different approach, \cite{Grady2019} calibrated the period--age relation by identifying LPV candidates in the LMC and Galactic star clusters with well-determined ages. The comparison with cluster Miras is particularly valuable as it provides the most direct constraint on the period--age relation. However, their results showed a significant offset of the period--age relation relative to those based on kinematics.

\begin{figure}[tb]
\begin{center}
\includegraphics[width=6in]{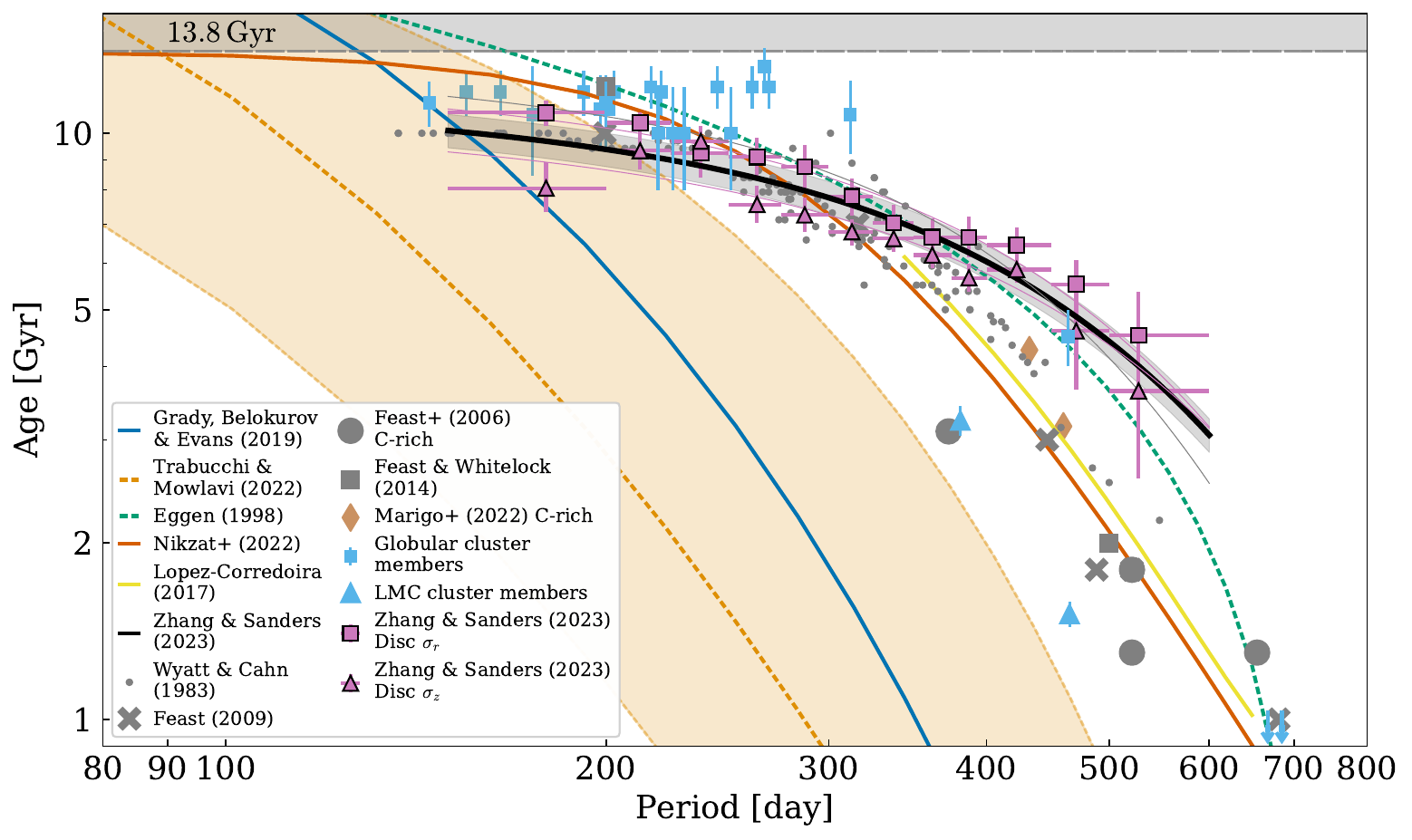}
\end{center}
\caption{Comparison of various period--age relation for Miras, available in the literature.  The models of \cite{Wyatt1983} are marked with small grey points, the green dashed line is from the model of \cite{Eggen1998}, the orange dashed line together with the orange region is from the model of \cite{Trabucchi2022}. The empirical relations from \cite{Lopez2017}, \cite{Grady2019}, and \cite{Nikzat2022} are marked with yellow, blue, and red solid lines, respectively. Large grey points represent period–age estimates for disc populations from \cite{Feast2006,Feast2009,Feast2014}.
Blue squares represent Miras in globular clusters \citep{Clement2001}, brown diamonds mark Miras in open clusters \citep{Marigo2022}, and blue triangles are Miras in LMC clusters.
Purple squares and triangles show the period–age measurements from \cite{ZhangPA2023}, and the black line represents their combined fit to these data together with globular cluster members. Figure courtesy of Hanyuan Zhang.}
\label{fig:PA}
\end{figure}

More recently, the period--age relation of O-rich Miras was calibrated by \cite{ZhangPA2023}, based on the Gaia DR3 LPV candidate catalog. The authors fitted dynamical models to astrometric data from Gaia to derive period–kinematic relations in the solar neighbourhood, which were then compared with literature age–velocity dispersion relations for red giants. The resulting relation agrees well with earlier studies, as well as when compared to cluster members of known age. 
Importantly, their selection of clusters was much more restrictive than in the work of \cite{Grady2019}, who used a broader and possibly contaminated sample of LMC cluster members.
The comparison of various observational and theoretical period--age relations is illustrated in \textbf{Figure~\ref{fig:PA}}. While there is a reasonable agreement between the majority of empirical relations, the recent theoretical calibrations tend to predict younger ages for a given period, demonstrating the need for further studies.

\section{APPLICATIONS OF THE LONG PERIOD VARIABLES}

The period--luminosity and period--age relations exhibited by LPVs are playing an increasingly significant role in measuring intergalactic and extragalactic distances, as well as analyzing the structure of the Milky Way. Although the period–luminosity relations are not as tight as those of Cepheid variables, the advantage is that pulsating red giant stars are far more common than Cepheids and they are present in all types of galaxies. Moreover, in the infrared range, stars at the tip of the AGB are brighter than Cepheids, which is particularly important in an era when large infrared space telescopes, such as the James Webb Space Telescope (JWST), are used to measure distances to supernova host galaxies, aiming to improve the precision of Hubble constant measurements.

Given their high luminosity and regular variability, Miras have traditionally been the primary focus among LPVs, and therefore will be the main topic of the following chapters. However, similar applications as distance indicators can also be extended to SRVs and even OSARGs, as these lower-amplitude LPVs also follow well-defined period--luminosity relations. Despite being often overlooked due to their less regular variability and greater observational challenges, they represent a promising and underexploited resource for stellar population studies, particularly in light of the capabilities offered by modern time-domain surveys. In particular, the potential of SRVs as distance indicators has recently been explored by \cite{Trabucchi2021srv}.

In addition to their role in distance measurements, the multiple period–luminosity relations of LPVs provide valuable diagnostics for the pulsation mechanisms of red giants and can be used to identify potential stellar companions, further underscoring their importance for stellar populations studies.

\begin{textbox}[]\section{HOW ARE DISTANCES ESTIMATED?}
The distance $d$ to a star can be determined by comparing its apparent magnitude $m$ and absolute magnitude $M$. The difference between these two quantities is known as the distance modulus $\mu$, given by the equation:
$$\mu = m - M = 5 \; \rm{log} \, d - 5$$
where the distance $d$ is expressed in parsecs and extinction is assumed to be negligible.
For pulsating stars, the absolute magnitude $M$ is estimated using the period--luminosity relation: $$M = a \; \rm{log} \, P +b$$
where $P$ is the pulsation period, while $a$ and $b$ are the coefficients of the relation. Since both the apparent magnitude $m$ and the pulsation period $P$ can be measured with high precision from observations, this method enables a highly accurate and direct determination of stellar distances. The relation may contain additional terms such as a metallicity dependence.
\subsection{What if extinction is non-negligible?}
This is often the case in studies of the Milky Way. The typical approach then is to use the Wesenheit index -- which is, by definition, reddening-free -- as a measure of stellar brightness, to avoid the complexities of extinction correction. However, recent studies have shown that the requirement of the Wesenheit index for a universal total-to-selective extinction ratio is not satisfied within our and other galaxies \citep{ZhangGreen2025}, making it unreliable for fully correcting extinction effects.\end{textbox}

\subsection{Distance indicators}
\label{sec:distance}

Mira variables have been widely used to determine distances within the Milky Way and to the Local Group galaxies, owing to their high luminosity and well-defined period--luminosity relations (see the sidebar titled HOW ARE DISTANCES ESTIMATED). Recent empirical studies \citep{Goldman2019,Bhardwaj2025} suggest that the period--luminosity relation for Miras is either independent of iron abundance or only weakly dependent on metallicity, however, the models of \cite{Fadeyev2025} and \cite{Trabucchi2025} indicate that there is a sensitivity of the period--luminosity relation to chemical composition. If the empirical findings were correct, this characteristic may provide Miras with a significant advantage as distance indicators compared to, for instance, classical Cepheids, whose luminosities are clearly influenced by chemical composition. The {\it K}-band is generally preferred for ground-based observations, as it minimizes the scatter of the period--luminosity relation and aligns with the peak spectral output of Miras. At the same time, observations at this wavelength also reduce the impact of circumstellar absorption, which is more severe at shorter wavelengths, and circumstellar emission, more significant at longer wavelengths \citep{Whitelock2012}. The scatter can be further reduced by using maximum-light magnitudes instead of mean-light values \citep{Bhardwaj2019}.

The LMC has served as the primary reference for calibrating the period--luminosity relations of Mira variables. Its well-determined distance, extensive long-term monitoring, and large population of well-characterised Miras have enabled the derivation of precise slopes and zero points, which are frequently adopted for Galactic and extragalactic applications. However, because stellar populations differ in age, metallicity, and chemical composition, such LMC-based calibrations may not be universally applicable without correction, but the extent to which these factors influence the period--luminosity relation is not sufficiently constrained \citep{Whitelock2008,Goldman2019,Sanders2023,Chen2024}. Alternatively, theoretical modelling has been proposed as a means of assessing and compensating for these effects \citep{Qin2018}.

High-precision parallax measurements obtained with Very Long Baseline Interferometry (VLBI) and related techniques provide direct distances to Mira variables, enabling independent calibration of their K-band period--luminosity relation \citep[e.g.,][]{Nakagawa2014}. Although such measurements are highly precise, they are available only for a limited number of objects, while parallaxes for AGB stars from Gaia are currently considered to have large, asymmetric, and underestimated uncertainties, particularly for highly extended and evolved objects such as Miras \citep{Andriantsaralaza2022}.

Mira variables are intrinsically luminous stars, allowing them to be observed even in densely populated and dust-obscured regions, thus making them a valuable independent estimate of the distance to the Galactic Center (GC), and have already been used for that purpose almost half a century ago by \cite{Glass1982} who arrived at a value of $9.2 \pm 0.6$ kpc, using 70 Mira variables. With the plethora of data from modern time-domain surveys,  numerous estimates have been derived: \cite{Groenewegen2005} combined OGLE-II, DENIS, and 2MASS catalogs to identify 2,691 Miras, deriving a GC distance of $8.6-9.0$ kpc ($\pm0.7$ kpc), depending on the assumed metallicity dependence of the K-band period--luminosity relation; \cite{Matsunaga2009} applied a method for simultaneous estimation of distance and extinction using $JHK_s$ period--luminosity relations to 1,364 Miras, obtaining a distance of $8.24 \pm 0.08$ (stat.) $\pm 0.42$ (syst.) kpc; \cite{Catchpole2016} combined $JHKL$ observations of 643 outer bulge Miras with 8,057 inner bulge Miras from OGLE, MACHO, and 2MASS, finding $8.9 \pm 0.4$ kpc; \cite{Qin2018} analysed 1,863 Miras from SAAO and MACHO, reporting a GC distance of $7.9 \pm 0.3 $ kpc; and most recently, \cite{Iwanek2023} used 65,981 OGLE Miras to map the Milky Way’s three-dimensional structure, deriving $7.66 \pm 0.01$ (stat.) $\pm 0.39$ (sys.) kpc. All these results demonstrate that the GC distance measurements using Miras are in good agreement with other methods (within $2\sigma$), confirming their reliability as distance indicators.

M33 is a Local Group galaxy that has been the subject of independent distance measurements using Mira stars. \cite{Yuan2018} were the first to calculate a distance modulus of $24.80 \pm 0.06$ mag, based on 1,265 stars observed in various near-infrared surveys. Subsequent determinations by \cite{Ou2023} ($24.67 \pm 0.06$ mag) and \cite{Konchady2024} ($24.629 \pm 0.046$ mag), based on multiband data from different surveys and larger Mira samples, are in good agreement with each other and with literature values obtained using other tracers.

\begin{marginnote}
\entry{Local Group}{a gravitationally bound collection of galaxies that includes the Milky Way, Andromeda, M33, and several dwarf galaxies, spanning roughly 3 Mpc in diameter.}
\end{marginnote}

Miras have also been used to measure distances in several other Local Group galaxies, including dwarf spheroidals: Leo I \citep{Menzies2010}, Fornax \citep{Whitelock2009}, Phoenix \citep{Menzies2008}, and Sculptor \citep{Menzies2011b}, as well as dwarf irregulars such as NGC 6822 \citep{Whitelock2013} and IC 1613 \citep{Menzies2015}, as part of a SAAO survey. The dwarf spheroidal Miras were found to follow the same period--luminosity relations as those in the LMC when bolometric luminosities are used.
In dwarf irregulars, C-rich Miras show the same behaviour, whereas O-rich Miras tend to be brighter than predicted by the period--luminosity relation, likely due to hot-bottom burning.
The estimated distances were in agreement with distances from other methods.

Mira distances to galaxies beyond the Local Group have so far been obtained for only three systems: NGC~5128, NGC~1559, and M101. The distance modulus to NGC~5128, the nearest giant elliptical galaxy, was first measured by \cite{Rejkuba2004} using near-infrared VLT observations of Miras and a K-band period--luminosity relation calibrated with Hipparcos parallaxes and the LMC Mira period--luminosity relation. The resulting distance modulus of $27.96 \pm 0.11$ mag is in excellent agreement with the Tip of the Red Giant Branch (TRGB) distance, and NGC~5128 remains the only giant elliptical with a Mira-based distance.

The other two galaxies are of particular interest as hosts of Type Ia supernovae, allowing Miras to calibrate SN Ia peak luminosities. In NGC~1559, \cite{Huang2020} derived $\mu = 31.41 \pm 0.05 \pm 0.06$ mag, later supported by \cite{Sanders2023} and \cite{Bhardwaj2025}, while in M101, \cite{Huang2024a} and \cite{Bhardwaj2025} obtained $\mu = 29.10 \pm 0.06$ mag, and $29.07 \pm 0.04$, respectively, consistent with Cepheid and TRGB distances. These applications illustrate how Miras contribute to addressing the Hubble tension (see Section~\ref{sec:h0}).

\subsection{Galactic structure and evolution}
\label{sec:structure}

Studies of the Milky Way structure are difficult because of the solar system’s position within the Galactic disk, where extinction along the line of sight, particularly toward the central regions, is high and variable. Nevertheless, the structure of the Galaxy has been investigated extensively using a variety of tracers, including neutral and molecular gas, star forming regions, red clump stars, and several classes of pulsating variables.
Pulsating stars are of particular importance, as their period–luminosity relations provide reliable distance estimates across different stellar populations. RR Lyrae stars, tracing old stellar populations, are typically employed for studies of the bulge \citep{Pietrukowicz2020}, while young classical Cepheids serve as tracers of the Galactic disk and spiral arms \citep{Skowron2025}.
Mira variables, spanning a wide age range, offer a complementary tool for examining both the Milky Way bulge and disk.
The advent of large-scale variability surveys (see Section~\ref{sec:data}) has greatly expanded known variable star samples, yielding thousands of such objects and enabling detailed three-dimensional mapping of the Milky Way and nearby systems.

\subsubsection{Milky Way bulge}
\label{sec:bulge}

The morphology of the Galactic bulge has long been debated. Early infrared surveys revealed its non-axisymmetric and a characteristic peanut-like shape, which was partly attributed to extinction effects. This morphology has also been interpreted as evidence of a boxy bulge, similar to structures observed in external galaxies and predicted by dynamical models.
More recent studies have suggested an X-shaped bulge, particularly from analyses of red clump stars and infrared imaging, though these results have been questioned due to potential artifacts in data processing and uncertainties in the red clump luminosity function. Importantly, the X-shape is not observed in other stellar populations, such as RR Lyrae or young main-sequence stars, with evidence instead supporting a boxy or peanut-shaped structure. This suggests a possible dependence of bulge morphology on stellar population and metallicity. For a general review of the three-dimensional structure, age, kinematics, and chemistry of the bulge we refer the reader to \cite{Zoccali2024}.

\begin{marginnote}
\entry{galactic bulge}{a central, dense region of a galaxy with complex morphology and kinematics, containing a mix of old and young stars}
\end{marginnote}

\begin{marginnote}
\entry{red clump}{core-helium-burning stars on the Hertzsprung–Russell diagram, appearing as a dense clump of similar luminosity and color}
\end{marginnote}

Mira-based studies of the Galactic bulge have spanned more than seven decades and are comprehensively reviewed by \cite{Catchpole2016}. Here, we focus on more recent works that have utilized variable star data from large-scale surveys.

\cite{Catchpole2016} combined their observations of 643 bulge O-rich Mira variables with  8057 Miras from OGLE \citep{Soszynski2013} and MACHO \citep{Bernhard2013}, to estimate distances and study bulge structure as a function of age, thanks to the period--age relation. They found that younger, long-period Miras ($\sim5$~Gyr or younger) trace a clear bar structure inclined to the line of sight, which may be part of an X-shaped structure, while older, short-period Miras show a more spheroidal distribution in the outer bulge and a distinct, possibly more complex, distribution in the inner bulge. \cite{Lopez2017} did a similar analysis based on the same dataset. The mean age of the Mira sample was estimated at $\sim9$~Gyr. They found only a single density peak along the studied lines of sight, in contrast to the double peak predicted by an X-shaped bulge, and showed that a boxy distribution provides a significantly better fit, excluding the X-shape at the $3.3 \sigma$ level. Given the comparable ages of Miras and red clumps, the Mira results cast doubt on whether the X-shaped morphology inferred from red clump studies reflects a genuine structural component of the Galaxy, and suggest that the X-shape may not be a universal feature of the bulge.

Another attempt on mapping the bulge and disc as a function of stellar age was done by \cite{Grady2020} with the use of data from the Gaia DR2 catalog of long-period variables \citep{Mowlawi2018}. The authors found that the bulge is well described by a triaxial boxy distribution. 
The oldest Miras ($\sim9–10$~Gyr) showed little bar-like morphology, while the younger populations traced a bar inclined by $\sim21^\circ$ to the Sun–Galactic centre line, with the youngest displaying a pronounced peanut shape and X morphology. These results are in line with the findings of \cite{Catchpole2016}. The age-morphology dependence of the boxy/peanut bulge was also postulated by \cite{Semczuk2022}.

\cite{Chrobakova2022} used recent OGLE-IV catalog of 65,981 Mira variables \citep{Iwanek2022} to analyse the shape of the bulge. They derived density maps in a region far from the plane (700 pc $\leq |Z| <$ 1500 pc), which were fitted with density models of a boxy bulge and an X-shaped bulge. Similarly to \cite{Lopez2017}, they found that the boxy model provides a much better fit to the data than the X-shaped model. However, the number of Miras was insufficient to perform the analysis in different age bins.

A different method was undertaken by \cite{Iwanek2023}, who used the same collection of Miras from the OGLE survey \citep{Iwanek2022}, but determined their distances based on mid-IR data from the WISE and Spitzer space telescopes. They derived mean luminosities in up to seven mid-IR bands and applied the mid-IR period--luminosity relations \citep{Iwanek2021}. This approach was motivated by the fact that observations at longer wavelengths reduce the influence of interstellar extinction and resulted in a high accuracy of determined distances.
The Mira distribution was then fitted with a 44-parameter bulge model \citep{Sormani2022}, accounting for distance uncertainties. The adopted model includes parameters describing an X-shaped structure, and the results provide independent evidence in favor of such a morphology.
The discrepancy with the results of \cite{Chrobakova2022}, despite using the same dataset, is attributed by \cite{Iwanek2023} to the simplified approach of Chrobakova et al., which neglects distance uncertainties and employs approximate extinction corrections and transformations from the I-band to the K-band mean magnitudes.

A novel approach was used by \cite{Hey2023}, who were the first to utilize the entire OGLE-III catalog of LPVs in the Galactic bulge \citep{Soszynski2013}, not just the Miras. The catalog contains 232,406 sources, of which 6,528 are Miras, 33,235 are SRVs, and 192,643 are OSARGs. The derived period–luminosity–amplitude relations, calibrated with the LMC distance, were used to measure distances with uncertainties of the order of $10-15\%$. The derived distance to the Galactic center, $8,108 \pm 106$ (stat) $\pm 93$ (sys) pc, is consistent with values reported in the literature. Combining their distance catalog with Gaia kinematics, the authors mapped the velocity field beyond the Galactic center and found evidence that the bar is both bisymmetric and aligned with the inner disk.

Another study that used stars other than Miras was conducted by \cite{Zhang2024}, who used low-amplitude LPVs (mainly OSARGs) from Gaia DR3 to study the kinematics and dynamics of the Galactic bar. The advantage of using low-amplitude stars, as opposed to high-amplitude Miras, is the much greater reliability of their radial velocity measurements. The authors were able to measure both the pattern speed of the Galactic bar and its dynamical length, and perform an analysis of the orbital structure of the bar.
\begin{marginnote}[]
\entry{nuclear stellar disk (NSD)}{a dense, flattened stellar structure located at the centers of galaxies, including the Milky Way, smaller in scale than the bulge.}
\end{marginnote}

It is also important to note that Mira variables have been used to study the formation history of the Milky Way bulge/bar by studying the age distribution of stars in the Nuclear Stellar Disk (NSD). Using a Mira variable period–age relation of \cite{ZhangPA2023}, \cite{Sanders2024} found that the NSD formed about 8~Gyr ago, implying a relatively early formation time for the Milky Way bar.

\subsubsection{Milky Way disk}

The structure of the disk of our Galaxy remains a subject of debate. While recent studies agree that the disk is warped and flared \cite[e.g.,][]{Skowron2019,Drimmel2025}, there is no consensus regarding details of its spiral structure, including the number of spiral arms, their pitch angles, and shapes. 
The structure of the disk of our Galaxy has been investigated with various pulsating variable stars, though studies have focused primarily on classical Cepheids due to their very young ages. Mira variables span a wide range of ages and are much more numerous, offering an alternative tracers of various Milky Way structures, including the disk.

\begin{marginnote}
\entry{galactic disk}{a flattened, rotating structure found in spiral and some irregular galaxies, characterized by a concentration of stars, gas, and dust, containing star-forming regions}
\end{marginnote}

In their study of LPVs – primarily O-rich Miras – in the Milky Way, \cite{Grady2019,Grady2020} employed data from the Catalina, ASAS-SN, and Gaia surveys, using the period--age relation as a tool to investigate stellar age gradients across the Galaxy. They showed that stars in the outer disc are systematically younger than those in the inner regions, with the transition from disc-like to halo populations occurring around 15 kpc. The authors also reported a subset of O-rich Miras apparently displaced from the disc to vertical distances of about 10 kpc. In addition, they found evidence that the stellar disc is relatively thick and compact at older ages, but becomes thinner and more radially extended at younger ages, consistent with an inside-out, upside-down formation scenario for the Milky Way.

The three-dimensional distribution of Miras in the Galactic disk has been investigated by \cite{Urago2020}. Their sample consisted of 108 long-period (i.e., the younger) Miras in the solar vicinity, with distances determined from mid-IR photometry from WISE. The comparison of their face-on distribution with literature spiral arm models indicates that they may indeed trace the spiral arm structure of the Milky Way. Their edge-on distribution shows that about half of the sample belongs to the thin disk, while the other half can be associated with the thick disk.

The limitations resulting from a small sample size in the study of \cite{Urago2020} were overcome with the availability of 65,981 Miras from OGLE-IV \citep{Iwanek2022}, utilized by \cite{Iwanek2023} to construct the first detailed three-dimensional map of the entire Galactic disk using Miras as tracers of the Galactic structure (as described in Section~\ref{sec:bulge}), without separating Miras into different age bins. They found that the Mira distribution reveals a flared disk and possibly a warp, though the evidence for warping is less robust than in the case of classical Cepheids, making direct comparisons between the degree of warping among tracers challenging. On the other hand, previous studies have suggested that the extent of the Galactic warp depends on the age of the stellar population, with older tracers exhibiting a stronger warp, indicating that the average warp inferred from Miras is expected to exceed that derived from Cepheids. 

An attempt to investigate the far side of the Milky Way disk was undertaken by \cite{Albarracin2025}, who identified 3,602 Miras in the VVV survey across the Milky Way bulge and far disk, constituting the largest catalog of variable stars on the far side of the Galactic disk. The majority of O-rich Miras were found in the bulge region beyond the Galactic center, with proper motions consistent with Galactic rotation. Period--age analysis indicated that younger Miras are located at larger Galactocentric radii, which is consistent with previous studies, supporting an inside-out formation scenario for the Galactic disk. The sample also traced the Galactic warp, in agreement with results from younger stellar populations. The derived rotation curve showed that Miras near the bulge follow kinematics consistent with RR Lyrae, while at larger distances they align with red clump stars and Cepheids. This reflects the dominance of shorter-period (older) Miras in the inner regions, while younger, longer-period stars are more prevalent at larger radii.

\subsubsection{Magellanic System}

There has been relatively little effort devoted to investigating the spatial distribution of LPVs in the Magellanic Clouds, largely due to the absence of sufficiently comprehensive datasets. The OGLE catalogs of LPVs, which are typically used as the primary reference sample thanks to their careful manual classification of variable stars, are based on the OGLE-III phase and therefore cover only the central regions of these galaxies.

Using the first data release from the Gaia mission, \cite{Deason2017} combined Gaia measurements with 2MASS and WISE photometry to identify a sample of red giants displaying variability signatures consistent with Mira candidates in the outskirts of the LMC.
They found that the spatial distribution of Mira candidates beyond approximately $10^\circ$ from the center of the LMC does not match the morphology of the gaseous component and only limited evidence is found for a population of intermediate-age Mira candidates within the gaseous bridge connecting the LMC and SMC. Additional concentrations of Mira candidates are detected in the Northern LMC stream and in regions extending eastward of the LMC, which are interpreted as SMC stars that were tidally stripped and subsequently accreted by the LMC during past dynamical interactions between the two galaxies.

\subsubsection{Dwarf galaxies in the Local Group}

Dwarf galaxies are the most common type of galaxy in the Universe. Dwarf galaxies in the Local Group are relatively nearby, come in many shapes, and cover a wide range of chemical compositions, and are therefore excellent targets for studying how different stellar populations influence galaxy evolution.

\begin{marginnote}
\entry{a dwarf galaxy}{a small galaxy containing roughly a thousand to several billion stars}
\end{marginnote}

One of the most effective ways to study how galaxies form and change over time is by reconstructing their star formation histories. Evolved stars, such as AGB stars and red supergiants, are especially useful for this purpose. Their late evolutionary stages allow us to trace stellar populations that formed over a very long period of time, from about 10~Myr to 10~Gyr ago. In addition, their high brightness, which is directly related to the mass they were born with, makes them valuable tools for probing a galaxy’s history.

LPVs, encompassing both AGB stars and red supergiants, exhibit pronounced variability at optical wavelengths, where amplitudes are comparatively large and therefore more readily detectable. Their distribution in luminosity can then be used to reconstruct past star formation. For this reason, a monitoring survey of 55 Local Group dwarf galaxies with the 2.5 m Isaac Newton Telescope in La Palma was undertaken by \cite{Saremi2020}, aimed at identifying all LPVs and determining the star formation historiess of these galaxies. In a series of publications, the authors have already analyzed LPVs in Andromeda I \citep{Saremi2021}, Andromeda VII \citep{Navabi2021}, Sagittarius Dwarf Irregular galaxy \citep{Parto2023}, IC~10 \citep{Gholami2023}, and Andromeda IX \citep{Abdollahi2023}, not only reconstructing the star formation historiess but also providing estimates of the total stellar masses of the galaxies.

\subsection{The Hubble constant}
\label{sec:h0}

The Hubble constant ($H_0$), which quantifies the present-day expansion rate of the Universe, remains one of the most debated parameters in modern cosmology. Measurements anchored in the early Universe, primarily from the Planck satellite’s analysis of the cosmic microwave background, yield a value of $H_0 = 67.4 \pm 0.5$ km s$^{-1}$ Mpc$^{-1}$ \citep{Planck2020}, while late-time, distance-ladder methods consistently favor higher values, with the SH0ES team reporting $H_0 = 73.17 \pm 0.86$ km s$^{-1}$ Mpc$^{-1}$ from Cepheid-based calibrations \citep{Riess2022}, and the CCHP team reporting $H_0 = 70.39 \pm 1.22$ (stat) $\pm 1.33$ (sys) $\pm± 0.70$ (SN), based on the TRGB calibration \citep{Freedman2025}. The $>5\sigma$ discrepancy between the Planck and SH0ES results has become widely known as the “Hubble tension” and represents the most statistically robust conflict in cosmology today. A recent effort to find a consensus value for the local determination of $H_0$, by combining all available datasets, yielded $H_0 = 73.50 \pm 0.81$ km s$^{-1}$ Mpc$^{-1}$, which differs by $7.1\sigma$ from the early-Universe value of $H_0$, further strengthening the tension \citep{Casertano2025}.

Traditional distance-ladder approaches rely heavily on Cepheids and Type Ia supernovae, with complementary constraints from the TRGB. However, alternative stellar candles such as Miras, J-region AGB stars, Type II supernovae, Surface Brightness Fluctuations, and the Baryonic Tully–Fisher Relation are increasingly important as independent checks, as they often yield similarly high late-time $H_0$ values, reinforcing the tension.

Mira variables offer a promising complement to Cepheids and TRGB in the distance ladder. Short-period ($P \leq 400$ d), O-rich Miras represent an older and widespread stellar population, extending distance measurements to galaxies lacking young Cepheids while also being brighter than TRGB stars. Their utility is enhanced by the growing availability of infrared observations, as Miras can be identified and characterized entirely in the near- and mid-infrared, where they are intrinsically luminous and less affected by dust. Long-period Miras ($P \geq 400$ d) are even more luminous and could, in principle, allow direct determinations of $H_0$ without relying on Type Ia supernovae, though their period--luminosity relations are less well constrained than those of Cepheids.

Period--luminosity relations for Miras have been established for decades (see Section \ref{sec:pl}), but renewed interest has been driven by the Hubble tension. Absolute calibrations rely on nearby geometric anchors, primarily the LMC with a 1.2\% distance precision \citep{Pietrzynski2019} and the water megamaser galaxy NGC 4258 with 1.4\% precision \citep{Reid2019}. The Mira period--luminosity relations in these systems have been shown to be consistent with their geometric distances \citep{Huang2018,Huang2020,Bhardwaj2025}, providing a robust first rung for the Mira distance ladder.

Building on these anchors, Miras have recently been used to calibrate SN Ia luminosities. In NGC~1559 (the host of SN~2005df), 115 O-rich Miras were identified in HST NIR data, yielding a distance modulus of $31.41 \pm 0.05$ (stat) $\pm 0.06$ (sys) mag \citep{Huang2020}. This made NGC~1559 the most distant galaxy studied with Miras to date (distance $19.1 \pm 1.1$ Mpc) and the first SN Ia host where the SN fiducial luminosity was found to be consistent with the Cepheid-based value, resulting in $H_0 = 73.3 \pm 4.0$ km s$^{-1}$ Mpc$^{-1}$. In a subsequent study, \cite{Sanders2023} used Gaia DR3 parallaxes to calibrate preliminary period–luminosity relations of O-rich Mira variables in the 2MASS $J$, $H$, and $K_s$ bands in the Milky Way. They then applied these relations to the NGC 1559 Mira sample \citep{Huang2020} to measure the distance to this galaxy and estimate the Hubble constant, although the large scatter in the period--luminosity relations, caused by uncertainties in both parallaxes and mean magnitudes, limits the precision of the resulting $H_0$ estimate. A joint analysis of the NGC 1559 and NGC 4258 Mira samples, anchored by the NGC 4258 water maser together with Milky Way and LMC Miras, yields a value of $H_0 = 73.7 \pm 4.4$ km s$^{-1}$ Mpc$^{-1}$, in a very good agreement with \cite{Huang2024a}.

However, a more accurate and precise calibration of period--luminosity relations can be obtained when using Mira variables in star clusters rather than Galactic field Miras, due to the availability of independent distances and the possibility of obtaining a more reliable mean parallax from numerous cluster members. \cite{Bhardwaj2025} identified 41 O-rich Miras in Galactic globular clusters and used their near-infrared and optical photometry combined with Gaia proper motions to derive period--luminosity relations with significantly reduced scatter. The accuracy of these relations was confirmed by determining a distance modulus to the LMC, $18.45 \pm 0.04$ mag, in agreement with the 1.2\% precise geometric distance. Using the LMC and NGC 4258 as independent anchors, the authors applied the relation to NGC 1559 and M101 and derived $H_0 = 73.06 \pm 2.67$ km s$^{-1}$ Mpc$^{-1}$, in agreement with previous studies based on Miras, but with a significantly lower uncertainty.

Recently, 288 O-rich Mira candidates were discovered in M101 (the host of SN 2011fe), with 211 short-period objects used to derive a distance modulus of $29.10 \pm 0.06$ mag \citep{Huang2024a}. Combining this with the NGC 1559 calibration gave $H_0 = 72.4 \pm 3.0$ km s$^{-1}$ Mpc$^{-1}$, thus confirming the previous local Hubble constant measurement at approximately 95\% confidence.

For a detailed overview of the Hubble tension, see \cite{DiValentino2024book}, and for an in-depth discussion of the role of Miras in addressing this issue, see the dedicated chapter by \cite{Huang2024book}.

\subsection{Possible planet indicators}
\label{sec:lsp}

Red giant and supergiant stars exhibit photometric variability driven by a range of phenomena, including stellar pulsations, supergranular convection in the envelopes, and the presence of circumstellar mater. However, one type of variability has not been definitively explained to this day. The so-called long secondary periods (LSPs) range from about 200 to over 1000 days, which is an order of magnitude longer than the pulsation (primary) periods in the same stars. The LSP phenomenon is observed in at least a third of OSARGs and SRVs, and it remains unclear whether it also occurs in Miras. The amplitudes of brightness variations associated with LSP are often greater than the amplitudes of stellar pulsations and can reach up to 1 magnitude in the {\it V}-band.

\begin{figure}[p]
\includegraphics[width=5.5in]{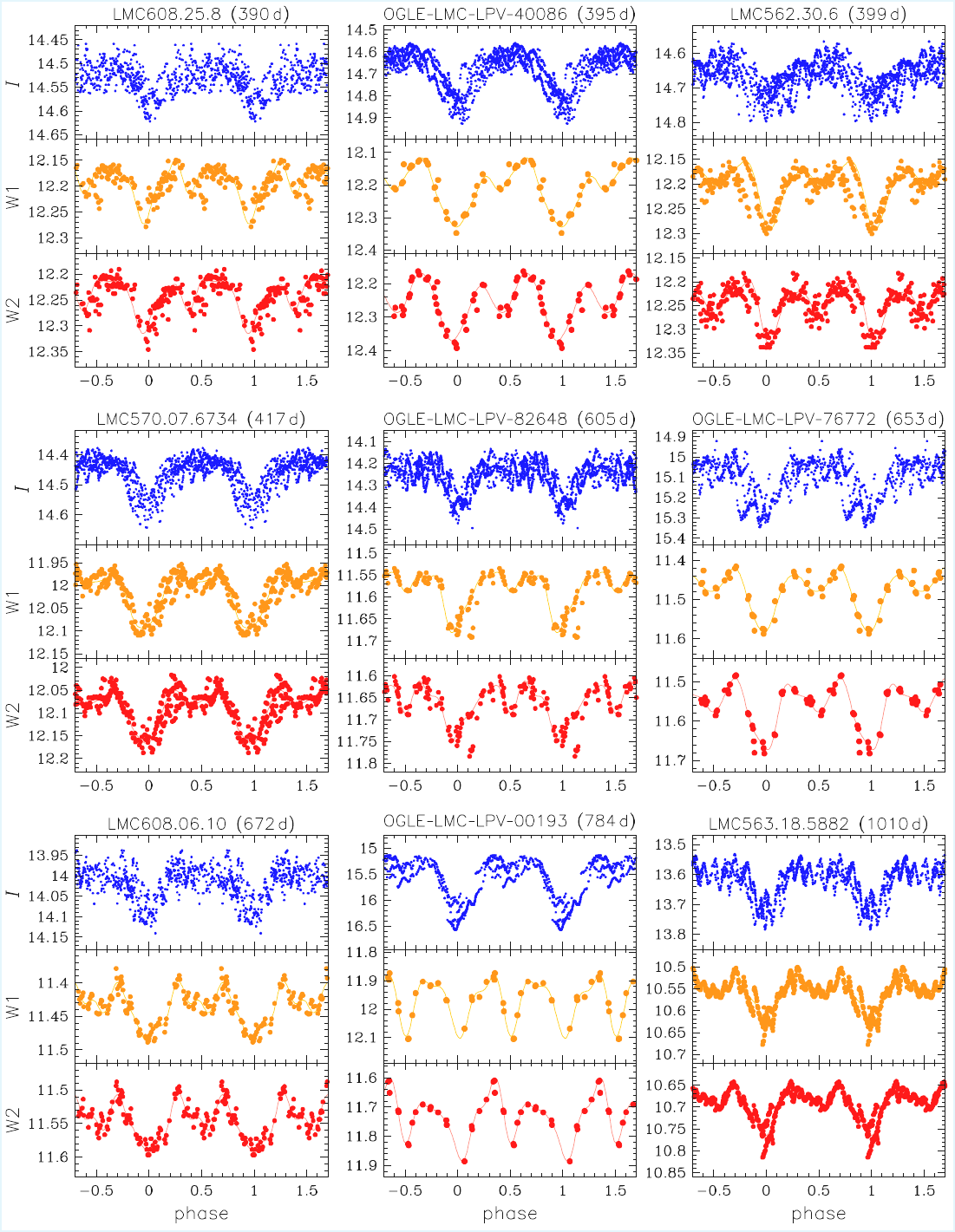}
\caption{Example optical and mid-infrared light curves of LSP variables. In each panel, blue, orange, and red points represent OGLE {\it I}-band photometry \citep{Soszynski2021}, NEOWISE W1, and NEOWISE W2 photometry \citep{Mainzer2014}, respectively. Note the secondary minima that are visible only in the infrared range. The LSPs are indicated in parentheses above each panel.}
\label{fig:LSP_lc}
\end{figure}

The optical light curves of LSP variables have been introduced in \textbf{Figure~\ref{fig:lsps}}, and as is readily notable, they have a distinctly different morphology from those produced by red giant pulsations. In visible light, the brightness variations during the LSP cycle can be divided into two roughly equal parts. In the first part, the light curve has a triangular shape -- the luminosity declines, reaches a minimum, and then rises again. During the second half of the LSP cycle, the star's brightness either remains nearly constant or increases gradually. This behaviour is not however the same in the infrared bands. \textbf{Figure~\ref{fig:LSP_lc}} shows a comparison of the optical (in blue) and mid-infrared (in orange and red) light curves of selected LSP variables, which are generally similar to those in the visible range but exhibit one notable difference: approximately half of all LSP variables show a secondary minimum in the mid-infrared, which is absent in the visual bands \citep{Soszynski2021}. This feature may be critical for understanding the nature of LSPs.

Infrared observations show that the LSP phenomenon is associated with the presence of circumstellar dust \citep{Wood2009,Pawlak2021}, and moreover, the distribution of this dust is not spherically symmetric. \citet{McDonald2019} linked the onset of LSPs in red giant stars with the onset of strong mass-loss due to dust-driven winds.

LSP variables follow a period--luminosity relation known as sequence D \citep{Wood1999}, which partially overlaps with and extends from sequence E toward higher luminosities and longer periods \citep{Soszynski2004a,Soszynski2004b}. Let us recall that sequence E is made up of close binary systems containing red giants, suggesting that LSP modulation may also be related to binarity. However, radial velocity measurements reveal a significant difference between the two groups: binary systems on sequence E exhibit velocity variation amplitudes on the order of several tens of km/s \citep{Nicholls2010}, whereas LSP variables show amplitudes an order of magnitude smaller -- typically between 2 and 7 km/s \citep{Hinkle2002,Wood2004,Nicholls2009}. If the LSP phenomenon is indeed caused by binarity, the companion to the red giant would have to be a substellar object (a brown dwarf) or a very low-mass star.

Over the years, various explanations for the origin of LSPs have been proposed, including radial and nonradial stellar pulsations, the turnover of giant convective cells, episodic dust ejections, and magnetic activity (see \citealt{Goldberg2024} for a critical review of the proposed hypotheses for LSPs). However, none of these mechanisms can fully account for all the observed properties of LSP variables. The binary hypothesis was proposed by \citet{Wood1999}. In this scenario, the red giant hosts a close-orbiting substellar or stellar companion embedded in a dusty cloud with a comet-like tail that periodically obscures the giant once per orbit. The binary model can explain the secondary minima observed in the infrared light curves of LSP variables \citep[Figure~\ref{fig:LSP_lc},][]{Soszynski2021}. The primary broad minima seen in both the visible and infrared light curves correspond to eclipses of the red giant by the dusty cloud surrounding the companion, while the secondary eclipses occur when the much cooler cloud passes behind the giant.

The binary scenario faces challenges in explaining LSPs. First, the radial velocity curves associated with LSPs are typically non-sinusoidal, suggesting eccentric orbits of potential red giant companions. However, the velocity curves of different LSP variables exhibit a strikingly similar shape, which is inconsistent with the expected uniform distribution of orbital periastron angles. Second, the phase shift between the light and radial velocity curves of LSP variables indicates that the densest part of the putative dusty cloud lies on the opposite side of the giant star relative to the low-mass companion \citep{Goldberg2024}. The first problem can be addressed by assuming that the non-sinusoidal velocity curve is not caused by the orbital eccentricity of a low-mass companion embedded in a dusty cloud, but rather by the Rossiter–McLaughlin effect, whereby the cloud obscures different regions of the giant's disk at various phases of the LSP cycle \citep{Soszynski2014}. The second problem was recently addressed by \citet{Decin2025}, who proposed a model where a low-mass companion on a close eccentric orbit captures dust near apastron, generating a trailing spiral structure with its densest region anti-phased relative to this companion.

The high incidence of LSPs among red giants (30--50\%) contrasts with the so-called ``brown-dwarf desert'' -- the observed scarcity of brown-dwarf companions around main-sequence stars which are the progenitors of red giants. This fact could be explained by assuming that the low-mass companion was originally a planet that accreted a substantial amount of material from the giant's envelope, evolving into a brown dwarf or even a low-mass main-sequence star. If this scenario is correct, LSP variables could serve as valuable tracers of extrasolar planetary systems in the Milky Way and beyond.

\subsection{Asteroseismology}

A wealth of observational data on LPVs from OGLE opened new opportunities for investigating the pulsational mechanisms in these variable stars, revealing the existence of multiple sequences in the period--luminosity plane.
While there is a general consensus that these sequences represent subsequent evolutionary stages in the life of an LPV, and that different sequences are associated with different LPV classes and pulsation modes (see Section~\ref{sec:pl}), the details of these relations are still not understood, with possible explanations including different pulsation mechanisms.
In addition to extensive ground-based catalogs of LPVs, the avaialbility of high-precision space-based photometry from Kepler and CoRoT has motivated intensified efforts to understand this phenomenon.

It has long been thought that the pulsations of evolved stars are driven by a partially ionized hydrogen layer via the $\kappa$ or $\gamma$ mechanism \citep{Keeley1970}, as observed in classical pulsators. At the same time, stars with convective envelopes typically exhibit oscillations caused by convective driving, seen in low-luminosity RGB stars \citep{Chaplin2013}, and theoretical models indicate that the amplitude of such oscillations grows rapidly with increasing stellar luminosity \citep{Christensen2001}.
The interplay between the above mechanisms is difficult to model \citep{Houdek2015}, but growing evidence suggests that solar-like oscillations are responsible for LPV variability.

Indeed, analyses of OGLE ground-based data have shown that OSARG pulsations are driven by stochastic excitation \citep{Dziembowski2010,Takayama2013,Xiong2025}, with the most recent study by \citet{Xiong2025} based on more than 70,000 OSARGs observed over 26 years.
An even longer dataset -- 38 years of observations of 39 SRVs analyzed by \citet{Cadmus2024} -- led to the same conclusion.
Also the asteroseismic analysis of Kepler data for LPVs carried out by \cite{Mosser2013} identified both radial and non radial solar-like oscillations, although non-radial pulsation modes were not detected by \cite{Hartig2014}. A later study of Kepler data by \cite{Yu2020}, however, confirmed that solar-like oscillations are the variability mechanism in SRVs.

Pulsation models suggest that red giants shift from stochastically excited to self-excited variability as luminosity increases, i.e., between the OSARG and SRV stars \citep{Xiong2018,Cunha2020,Trabucchi2025}, which is consistent with the Kepler data analysis \citep{Banyai2013}. Additional support comes from hydrodynamical simulations of AGB convective envelopes, which are able to reproduce the observed variability  \citep{Freytag2017,Trabucchi2021,Ahmad2023}.

\section{THE FUTURE} 

Extensive datasets on LPVs have already been used to study these stars, yet much of this information remains underexploited, continuing to serve as an invaluable resource for testing stellar evolution and pulsation models. Advances in computational techniques and a deeper understanding of the physical mechanisms driving LPV behavior have already led to progress in theoretical studies, but several issues remain to be addressed, including the complex interplay between convection and pulsation or the mechanism behind the long secondary periods.

Upcoming datasets will enable increasingly detailed studies, improving our understanding of the late stages of stellar evolution, the nature of LPV pulsations, and their contribution to the structure and evolution of the Milky Way and other galaxies. Future releases of the OGLE-IV catalog, covering the entirety of the Magellanic Clouds and the Milky Way disk, will continue to provide high-quality reference samples that facilitate analyses of LPV distributions and their dependence on metallicity. These releases will include 30-year light curves for hundreds of thousands of LPVs, allowing investigations not only of exceptionally long periods but also of period changes and mode switching, offering direct insight into variations driven by stellar evolution.

The upcoming Gaia DR4 release will cover $\sim5.5$ years and DR5 is planned to cover the full $\sim10.5$ years of the mission (versus $\sim3$ years in DR3), providing all time-series data for photometry, astrometry and spectroscopy.
The increased number of observations will enable better classification of LPVs as well as more accurate period estimation. Precise distances and proper motions will enable robust studies of LPV spatial distributions, while spectroscopic data will provide key stellar parameters and radial velocities.

Over the next decade, the Vera C. Rubin Observatory’s LSST will produce long-term, high-cadence, multi-band light curves for millions of LPVs, substantially expanding the known population of these stars and allowing detailed characterization of their variability and demographics. With its deep magnitude limit, reaching about 23.5 mag in the $r$-band, LSST will also enable the discovery and characterization of LPVs in more distant galaxies, extending variability studies beyond the Local Group.

The 4-metre Multi-Object Spectroscopic Telescope (4MOST) -- a wide-field spectroscopic survey instrument on the VISTA telescope, will supply large-scale spectroscopic data, including accurate radial velocities, chemical abundances, and fundamental stellar parameters. In particular, the "One Thousand and One Magellanic Fields" (1001MC) 4MOST survey will target approximately 25,000 O-rich AGB stars and 9,000 carbon stars in the Magellanic Clouds, and is expected to include thousands of pulsating red giants already classified as LPVs by OGLE. The resulting radial velocity measurements will help constrain the fraction of LSP variables, while the derived chemical abundances will enable detailed investigations of the chemical characteristics of LPVs.

Together, new observational datasets will be critical for increasing LPV samples, mapping LPV populations, probing their kinematics, understanding their role in Galactic chemical evolution, and providing stringent tests for stellar evolution and pulsation models.

Until now, only Miras have been widely used as distance indicators, but there is growing interest in using SRVs and OSARGs, which also follow period–luminosity relations. Expanding the set of LPVs used for galactic structure studies could increase available data by an order of magnitude and enable investigations of LPV distributions across different evolutionary stages. Using Miras, and potentially other LPVs, as standard candles requires careful calibration of their period–luminosity relations and consideration of their sensitivity to chemical composition, which will become increasingly feasible with the forthcoming observational data, especially with the availability of high-quality data from space telescopes such as JWST and Roman, and data from extensive time-domain and spectroscopic surveys from the ground. Together, these efforts will enable a more complete understanding of LPVs as both tracers and drivers of stellar and galactic evolution.

\begin{summary}[SUMMARY POINTS]
\begin{enumerate}
\item  LPVs are red giant and supergiant stars exhibiting periodic or quasi-periodic variations in brightness caused by pulsations, over timescales ranging from days to several years. They are typically found on the upper RGB or AGB and cover a wide range of stellar masses, from sub-solar to several tens of solar masses. The amplitudes of their variability can span from millimagnitudes, detectable only with precise photometry, to several magnitudes, easily observable with ground-based telescopes. LPV light curves  often display a combination of regular and irregular components.

\item The contemporary classification of LPVs distinguishes three classes: Miras, SRVs, and OSARGs. Miras pulsate in the fundamental mode with large variability amplitudes, have longest pulsations periods, and mostly regular light variations. SRVs are characterized by lower amplitudes, shorter periods, and less regular light variations, and can display two or three excited pulsation modes, usually including the fundamental and first overtone. OSARGs exhibit low-amplitude multi-periodic light variations as a result of simultaneous pulsation in radial and non-radial modes.

\item Long secondary periods in red giants are long-term brightness variations observed in a large fraction of OSARGs and SRVs. The leading explanation of this phenomenon involves a low-mass companion embedded in a dusty cloud, supported by their distinctive optical and infrared light curves, making LSP variables potential tracers of evolved planetary systems.

\item LPVs obey period--luminosity relations, where brighter stars pulsate with longer periods. The plethora of data from large-scale time-domain surveys revealed multiple sequences on the period--luminosity plane, associated with Miras, SRVs, OSARGs, LSPs and ellipsoidal variables making the period--luminosity diagram a powerful tool to probe the pulsation and evolution of red giant stars.

\item LPVs follow a period--age relation, with younger stars pulsating at longer periods. However, structural changes during the AGB phase introduce significant scatter, limiting their precision as age indicators unless carefully modeled. Comparisons of observational and theoretical period--age relations show reasonable agreement among empirical results, though recent models tend to predict younger ages for a given period.

\item  Mira variables are valuable distance indicators due to their high luminosity and well-defined period--luminosity relations, which are relatively insensitive to metallicity. Miras have been used to measure distances within the Milky Way, including the Galactic Center, and to Local Group galaxies, with results consistent with other tracers. More recently, their use has extended beyond the Local Group, including supernova-host galaxies, where they aid in calibrating Type Ia supernova luminosities and addressing the Hubble tension.

\item Miras serve as tracers of Galactic and extragalactic structure, providing distances, ages, and star formation histories that reveal the three-dimensional morphology, kinematics, and evolution of the Milky Way and Local Group galaxies.

\end{enumerate}
\end{summary}

\section*{DISCLOSURE STATEMENT}
The authors are not aware of any affiliations, memberships, funding, or financial holdings that might be perceived as affecting the objectivity of this review. 

\section*{ACKNOWLEDGMENTS}
The authors thank Jan Skowron for \textbf{Figure~\ref{fig:classification}} and Hanyuan Zhang for \textbf{Figure~\ref{fig:PA}}. DMS acknowledges support from the European Union (ERC, LSP-MIST, 101040160). Views and opinions expressed are however those of the authors only and do not necessarily reflect those of the European Union or the European Research Council. Neither the European Union nor the granting authority can be held responsible for them. This work has been funded by the National Science Centre, Poland, through grant 2022/45/B/ST9/00243 to IS. For the purpose of Open Access, the author has applied a CC-BY public copyright license to any Author Accepted Manuscript (AAM) version arising from this submission.

\bibliography{paper.bib}{}
\bibliographystyle{ar-style2.bst}

\end{document}